\documentclass[apj,numberedappendix]{emulateapj}
\usepackage{geometry} 
\usepackage{natbib}
\usepackage{graphics}
\usepackage{graphicx}
\usepackage{amssymb}
\usepackage{amsmath}
\usepackage{epstopdf}
\usepackage{epsfig}
\usepackage[usenames,dvipsnames]{color}
\usepackage[normalem]{ulem}

\newcommand{\beqn}{\begin{equation}}
\newcommand{\eeqn}{\end{equation}}

\newcommand{\ie}{\emph{i.e.}\ }
\newcommand{\cf}{\emph{c.f.}\ }
\newcommand{\eg}{\emph{e.g.}\ }

\newcommand{\kms}{km~s$^{-1}$}

\newcommand{\vlsr}{$v_{_\mathrm{LSR}}$}
\newcommand{\dtan}{$d_\mathrm{tan}$}
\newcommand{\dml}{$d_{_\mathrm{ML}}$}
\newcommand{\dbar}{$\bar{d}$}
\newcommand{\dsun}{$d_{_\sun}$}
\newcommand{\spitzer}{{\it Spitzer}}

\newcommand{\HI}{\ion{H}{1}}
\newcommand{\HII}{\ion{H}{2}}
\newcommand{\htwo}{H$_2$}
\newcommand{\hcop}{HCO$^+$}
\newcommand{\nnhp}{N$_2$H$^+$}
\newcommand{\nhhh}{NH$_3$}
\newcommand{\thco}{$^{13}$CO}
\newcommand{\twco}{$^{12}$CO}

\newcommand{\ffore}{$f_\mathrm{fore}$}

\newcommand{\kband}{\langle\kappa\rangle_{_\mathrm{band}}}
\newcommand{\tauband}{\langle\tau\rangle_{_\mathrm{band}}}
\newcommand{\pchoose}{$P_{_\mathrm{ML}}$}
\newcommand{\lbd}{($\ell,b$,\dsun)}
\newcommand{\rphiz}{($R_\mathrm{gal},\phi,z$)}

\shorttitle{The Bolocam Galactic Plane Survey. VIII.}
\shortauthors{Ellsworth-Bowers et al.}

\slugcomment{Accepted Version: \today}

\begin{document}
\bibliographystyle{apj}

\title{The Bolocam Galactic Plane Survey. VIII.  A Mid-Infrared Kinematic Distance Discrimination Method}

\author{Timothy P. Ellsworth-Bowers\altaffilmark{1,2}, Jason Glenn\altaffilmark{1}, Erik Rosolowsky\altaffilmark{3}, Steven Mairs\altaffilmark{4}, Neal J. Evans II\altaffilmark{5}, Cara Battersby\altaffilmark{1}, Adam Ginsburg\altaffilmark{1}, Yancy L. Shirley\altaffilmark{6,7}, John Bally\altaffilmark{1}}
\altaffiltext{1}{CASA, University of Colorado, UCB 389, University of Colorado, Boulder, CO 80309, USA}
\altaffiltext{2}{email: \texttt{timothy.ellsworthbowers@colorado.edu}}
\altaffiltext{3}{Department of Physics and Astronomy, University of British Columbia Okanagan, 3333 University Way, Kelowna BC, V1V 1V7, Canada}
\altaffiltext{4}{Department of Physics and Astronomy, University of Victoria, 3800 Finnerty Road, Victoria, BC V8P 1A1, Canada}
\altaffiltext{5}{Department of Astronomy, University of Texas, 1 University Station C1400, Austin, TX 78712, USA}
\altaffiltext{6}{Steward Observatory, University of Arizona, 933 North Cherry Avenue, Tucson, AZ 85721, USA}
\altaffiltext{7}{Adjunct Astronomer at the National Radio Astronomy Observatory}

\begin{abstract}

We present a new distance estimation method for dust-continuum-identified molecular cloud clumps.  Recent (sub-)millimeter Galactic plane surveys have cataloged tens of thousands of these objects, plausible precursors to stellar clusters, but detailed study of their physical properties requires robust distance determinations.  We derive Bayesian distance probability density functions (DPDFs) for 770 objects from the Bolocam Galactic Plane Survey in the Galactic longitude range $7\fdg5 \leq \ell \leq 65\degr$.  The DPDF formalism is based on kinematic distances, and uses any number of external data sets to place prior distance probabilities to resolve the kinematic distance ambiguity (KDA) for objects in the inner Galaxy.  We present here priors related to the mid-infrared absorption of dust in dense molecular regions and the distribution of molecular gas in the Galactic disk.  By assuming a numerical model of Galactic mid-infrared emission and simple radiative transfer, we match the morphology of (sub-)millimeter thermal dust emission with mid-infrared absorption to compute a prior DPDF for distance discrimination.  Selecting objects first from (sub-)millimeter source catalogs avoids a bias towards the darkest infrared dark clouds (IRDCs) and extends the range of heliocentric distance probed by mid-infrared extinction and includes lower-contrast sources.  We derive well-constrained KDA resolutions for 618 molecular cloud clumps, with approximately 15\% placed at or beyond the tangent distance.  Objects with mid-infrared contrast sufficient to be cataloged as IRDCs are generally placed at the near kinematic distance.  Distance comparisons with Galactic Ring Survey KDA resolutions yield a 92\% agreement.  A face-on view of the Milky Way using resolved distances reveals sections of the Sagittarius and Scutum-Centaurus Arms.  This KDA-resolution method for large catalogs of sources through the combination of (sub-)millimeter and mid-infrared observations of molecular cloud clumps is generally applicable to other dust-continuum Galactic plane surveys.

\end{abstract}

\keywords{Galaxy: kinematics and dynamics, structure  -- infrared: ISM -- ISM: clouds, dust -- stars: formation}


\section{Introduction}\label{sec:intro}

Recent (sub-)millimeter surveys of the Galactic plane (ATLASGAL, \citealp{Schuller:2009}; Hi-GAL, \citealp{Molinari:2010}; BGPS, \citealp{Aguirre:2011}) have detected tens of thousands of molecular cloud cores and clumps in thermal dust emission.  As plausible precursors to stellar clusters, OB associations, or smaller stellar groups, molecular cloud clumps can yield clues about the formation of massive stars \citep{McKee:2007}.  The masses and temperature profiles of these objects are key to unraveling this process.  Recent work has sought to measure these quantities \citep{Russeil:2011,Eden:2012}, but a robust and comprehensive tally does not yet exist.

Derivation of masses for molecular cloud clumps from dust continuum data requires an estimate of the heliocentric distance to each object and the temperature of the emitting dust.  Analysis of {\it Herschel} Hi-GAL data is beginning to yield temperature maps of the Galactic plane \citep{Peretto:2010a,Battersby:2011}.  While a detailed understanding of the interplay between dust temperature and the environment and evolution of molecular cloud clumps is important, variations in the assumed dust temperature by a factor of two only produce a factor of a few difference in the mass derived from (sub-)millimeter observations.  In contrast, the derived mass of a molecular cloud clump is proportional to the square of its heliocentric distance;  accurate distance estimates play a far larger role in the mass calculation.  Recent studies of isolated regions, with well-determined distances, such as Perseus and Ophiuchus \citep{Ridge:2006a,Enoch:2006,Rosolowsky:2008}, have unveiled many properties of molecular cloud cores in recent years \citep{Enoch:2007,Schnee:2010}.  To gain similar insight into the larger molecular cloud clumps seen spread throughout the Galactic plane, a robust method for distance determinations for large data sets is required, because the distances to most clumps are subject to the kinematic distance ambiguity (KDA).

The most straightforward method for estimating the heliocentric distance (\dsun) to a molecular cloud clump is to project its observed line-of-sight velocity (\vlsr), derived from molecular line Doppler shifts, onto a Galactic rotation curve.  These kinematic distances are generally unique for the outer Galaxy, but inner Galaxy sources are subject to the KDA, a projection effect of the orbital motion for objects within the Solar Circle ($R_0$).   A line of sight intersecting a circular orbit at Galactocentric radius $R_\mathrm{gal} < R_0$ crosses that orbit twice, each with different spatial velocities but both with the same \vlsr.  Various techniques have been suggested for resolving the KDA (21-cm \HI\ absorption: \citealp{Anderson:2009a}, \citealp{RomanDuval:2009}; the presence of mid-infrared dark clouds: \citealp{Rathborne:2006}, \citealp{Peretto:2009}; H$_2$CO absorption: \citealp{Sewilo:2004}; and near-infrared extinction: \citealp{Marshall:2009}, \citealp{Foster:2012}); this paper presents a method based on comparing mid-infrared extinction with (sub-)millimeter emission.

Appearing as dark absorption features against a bright mid-infrared background, infrared dark clouds (IRDCs) offer a practicable means for resolving the KDA.  IRDCs are most striking against the broad, diffuse Galactic emission near $\lambda = 8$~\micron\ \citep{Perault:1996,Simon:2006}, although they may be detected in absorption against background stars at other infrared wavelengths \citep[\cf][]{Foster:2012}.  Studies of IRDCs at (sub-)millimeter wavelengths reveal that they are dense molecular cloud clumps \citep{Johnstone:2003,Rathborne:2006,Battersby:2010,Battersby:2011}.  As extinction features, IRDCs must lie in front of enough mid-infrared emission to be visible.  It is possible to {\it a priori} assign the near kinematic distance for the darkest clouds \citep[\eg][]{Butler:2009,Peretto:2009}, but recent work by \citet{Battersby:2011} has shown that molecular cloud clumps may be visible as slight intensity decrements in the mid-infrared at the far kinematic distance despite not being dark enough to be cataloged as IRDCs.  To encompass this second set of objects, we classify all dust-continuum-identified molecular cloud clumps with mid-infrared intensity decrements of any amount as Eight-Micron Absorption Features (EMAFs), whether catalogued as either an IRDC or not.  These constitute a generalized collection of cold molecular cloud clumps identified first by dust-continuum emission and then checked for infrared absorption.  The EMAF definition excludes objects extensively undergoing the later stages of star formation or that are exposed to strong ultraviolet radiation, as both processes excite PAH emission near $\lambda = 8$~\micron, rendering invisible any absorption.  Investigating the mid-infrared properties of molecular cloud clumps based on this classification avoids a bias toward the darkest, nearby IRDCs.

This paper presents a quantitative distance estimation technique for molecular cloud clumps based on Bayes' Theorem.  A distance probability density function (DPDF) is computed using a distance likelihood derived from kinematic information (observed \vlsr) and prior probabilities, based on ancillary data sets, that are applied in an effort to resolve the KDA.  We present here two such priors.  The first involves the comparison between observed mid-infrared absorption and millimeter emission of individual molecular cloud clumps, and the second is based on the Galactic-scale distribution of molecular gas.  In addition to those described here, any number of additional priors may be applied to constrain the distance estimate.

The apparent optical depth of an EMAF calculated na\"ively from mid-infrared images is likely less than the true value due to diffuse 8-\micron\ emission lying between the cloud and the observer.  By parameterizing the amount of total mid-infrared emission along a line of sight lying in front of a molecular cloud clump as the ``foreground fraction'' (\ffore), simple radiative transfer arguments may be used to derive the true optical depth.  The recent numerical Galactic infrared emission model of \citet{Robitaille:2012} offers an estimate of \ffore\ as a function of \dsun\ in the Galactic plane.  The maximum likelihood distance to a molecular cloud clump may be derived by comparing the optical depth calculated from (sub-)millimeter thermal dust continuum data with the absorption optical depth derived from the mid-infrared images and \ffore(\dsun).  This comparison generates a DPDF that takes into account Galactic-scale conditions along a given line of sight, including spiral structure.  A DPDF derived in this manner contrasts the widely-used ``step-function'' method whereby a molecular cloud clump is automatically assigned the near kinematic distance upon association with a catalogued IRDC.

The methodology presented here is valid only for molecular cloud clumps that exhibit mid-infrared absorption, and therefore is but one means for distance discrimination for large catalogs of dust-continuum-identified objects.  We present an automated means for deriving Bayesian DPDFs for mid-infrared dark molecular cloud clumps detected by the Bolocam Galactic Plane Survey, but this method is applicable to all (sub-)millimeter Galactic plane surveys.

This paper is organized as follows.  Section \ref{sec:data} describes the data sets used.  The DPDF formalism is described in Section \ref{sec:dpdf}.  Section \ref{sec:morph} outlines the generation of prior DPDFs for EMAFs.  Results from the Bayesian DPDFs are presented in Section \ref{sec:results}. Implications of this work are discussed in Section \ref{sec:discuss}, and conclusions are presented in Section \ref{sec:conclusion}.


\section{Data Sets}\label{sec:data}

\begin{deluxetable*}{ccccccc}
  \tablecolumns{7}
  \tablewidth{0pt}
  \tabletypesize{\small}
  \tablecaption{Spectroscopic Follow-up Observations of Inner Galaxy BGPS Sources\label{table:molec}}
  \tablehead{
    \colhead{Species} & \colhead{Transition} & \colhead{$\nu$} & 
    \colhead{Resol.\tablenotemark{a}} & 
    \colhead{$n_{eff}$\tablenotemark{b}}  &
    \colhead{N$_{_\mathrm{BGPS}}$\tablenotemark{c}}  & \colhead{Ref.} \\
    \colhead{} & \colhead{} & \colhead{(GHz)} & 
    \colhead{(\arcsec)} & \colhead{(cm$^{-3}$)} & 
    \colhead{} }
  \startdata
  \hcop    &   $J=3-2$  &  267.6   & 28  & $ 10^4$  & 6194  & 1 \\
  \nnhp    &   $J=3-2$  &  279.5   & 27  & $ 10^4$  & 6194  & 1 \\
  CS       &   $J=2-1$  &  97.98   & 64  & $ 5\times10^3$  & 553  & 2 \\
  \nhhh    &  (1,1)     &  23.69   & 31  & $ 10^3$ & 631  & 3
  \enddata
  \tablenotetext{a}{Beam FWHM}
  \tablenotetext{b}{Approximate effective density for line excitation at $T = 20$~K \citep{Evans:1999}}
  \tablenotetext{c}{Number of unique BGPS sources observed in this line}
  \tablerefs{(1) \citet{Shirley:2013}; (2) Y. Shirley (2012, private communication); (3) \citet{Dunham:2011c}}
\end{deluxetable*}

\subsection{The Bolocam Galactic Plane Survey}\label{data:bgps}

The Bolocam Galactic Plane Survey \citep[BGPS;][]{Aguirre:2011,Ginsburg:2013} is a $\lambda = 1.1$ mm continuum survey covering 170 deg$^2$ at 33\arcsec\ resolution.  The BGPS was observed with the Bolocam instrument at the Caltech Submillimeter Observatory (CSO) on Mauna Kea.  It is one of the first large-scale blind surveys of the Galactic Plane in this region of the spectrum, covering $-10\degr \leq \ell \leq 90\degr$ with at least $|b| \leq 0\fdg5$, plus selected regions in the outer Galaxy.  For a map of BGPS V1.0 coverage and details about observation methods and the data reduction pipeline, see \citet[][hereafter A11]{Aguirre:2011}.

From the BGPS V1.0 images, 8,358 millimeter dust-continuum sources were identified using a custom extraction pipeline.  The BGPS catalog (Bolocat) contains source positions, sizes, and flux densities extracted in various apertures, among other quantities (see \citealp{Rosolowsky:2010} for complete details).  BGPS V1.0 pipeline products, including image mosaics and the catalog, are publicly available\footnote{Available through IPAC at\\ \texttt{http://irsa.ipac.caltech.edu/data/BOLOCAM\_GPS}}.  For this work, we utilized the flux densities measured in a 40\arcsec\ top-hat aperture, which has the same solid angle as the BGPS 33\arcsec\ FWHM Gaussian beam ($\Omega = 2.9 \times 10^{-8}$~sr), in addition to the map data.  A flux calibration multiplier of $1.5 \pm 0.15$ was applied to both Bolocat and the image mosaics to correct a V1.0 pipeline error (see A11 and \citealp{Ginsburg:2013} for a full discussion).

The BGPS data pipeline removes atmospheric signal using a principle component analysis technique that discards time-stream signals correlated spatially across the bolometer array.  This effectively acts as an angular filter, attenuating angular scales comparable to or larger than the array field of view (see A11, their Fig.~15).  The implication is that the BGPS is not sensitive to scales larger than 6\arcmin.  The effective angular size range of detected BGPS sources therefore corresponds to anything from molecular cloud cores up to entire clouds depending on the heliocentric distance \citep{Dunham:2011c}.  In this work we refer to BGPS objects as ``molecular cloud clumps'' for simplicity, but recognize that distant sources are likely larger structures.

\subsection{Spectroscopic Follow-Up of BGPS Sources}\label{data:spec}

Several spectroscopic follow-up programs have been conducted to observe BGPS sources in a variety of molecular emission lines that trace the dense gas associated with molecular cloud clumps.  These surveys provide both kinematic and chemical information, and are typically beam-matched to the BGPS to facilitate comparison to the dust-continuum data.  From these observations, a line-of-sight velocity (\vlsr) was successfully fitted for each of more than 3,500 detected sources.  A summary of spectroscopic programs is presented in Table \ref{table:molec}.

In a pilot study \citep{Schlingman:2011} and complete survey \citep{Shirley:2013}, all 6,194 Bolocat objects at $\ell \geq 7\fdg5$ were observed using the Heinrich Hertz Submillimeter Telescope (HHT) on Mt.~Graham, Arizona.  These studies simultaneously observed the $J$=3$-$2 rotational transitions of \hcop\ ($\nu = 267.6$~GHz) and \nnhp\ ($\nu = 279.5$~GHz).  Because these molecular transitions trace fairly dense gas ($n_\mathrm{eff} \approx 10^{4}$~cm$^{-3}$)\footnote{The effective density required to produce line emission with a brightness temperature of 1~K; may be up to several orders of magnitude smaller than the critical density \citep{Evans:1999}.}, the line-of-sight confusion seen in CO studies is largely absent.  In fact, \citeauthor{Shirley:2013} find only 2.5\% of \hcop\ detections have multiple velocity components.  These objects, likely an overlap of two or more molecular cloud clumps along the line of sight, are not used in this study.  Detectability in \hcop\ is a strong function of millimeter flux density, and the detection rate for the full HHT survey was $\approx 50$\% (see \citealp{Shirley:2013} for full details).  Velocity fits to \hcop\ spectra constitute the bulk of the kinematic data used in this study (\nnhp\ spectra were not used because the complex hyperfine structure of its transitions makes it difficult to fit \vlsr).

As a companion to the HHT observations, a subset of 555 BGPS sources were observed in the $J$=2$-$1 rotational transition of CS ($\nu = 97.98$~GHz) using the Arizona Radio Observatory 12m telescope on Kitt Peak (Y. Shirley 2012, private communication; see \citealp{Bally:2010b}).  This subset was confined to $29\degr \leq \ell \leq 31\degr$, a region with a high density of sources looking toward the Molecular Ring and the end of the long Galactic bar.  This transition of CS traces lower density gas ($n_\mathrm{eff} \approx 5 \times 10^{3}$~cm$^{-3}$) than \hcop($3-2$), and was detected in 45\% of sources not detected by the HHT survey in this region.

Seeking to characterize the physical properties of BGPS sources, \citet{Dunham:2011c} used the Robert F. Byrd Green Bank Telescope to observe the lowest inversion transition lines of \nhhh\ near 24 GHz.  They observed 631 BGPS sources in the inner Galaxy.  The \nhhh\ (1,1) inversion is the strongest ammonia transition at the cold temperatures of BGPS sources ($T \approx 20$~K), and we used this transition exclusively for the \nhhh\ velocity fits.

\subsection{The \spitzer\ GLIMPSE Survey}\label{data:glimpse}

The \spitzer\ GLIMPSE survey \citep{Benjamin:2003,Churchwell:2009} was used to identify mid-infrared extinction features associated with BGPS detected sources.  The GLIMPSE survey area completely encompasses the BGPS for $|b| \leq 1\fdg0$ and $\ell \leq 65$\degr (there are several sections of the BGPS that flare out to $|b| \leq 1\fdg5$, see A11).  We used the V3.5 IRAC Band~4 mosaics\footnote{Data product manual:\\ \texttt{http://irsa.ipac.caltech.edu/data/SPITZER/GLIMPSE/}\\\texttt{doc/glimpse1\_dataprod\_v2.0.pdf}} ($\lambda_c = 7.9$~\micron) to identify absorption features.  Point sources (stars) identified in the Band~1 mosaics ($\lambda_c = 3.6$~\micron) were removed from the Band~4 images to accentuate diffuse emission (see \S \ref{sec:ubc_proc}).  Stars were modeled as Gaussian peaks since the mosaicing process from individual IRAC frames produces a spatially variable PSF, hampering star-subtraction.  The Band~4 mosaics have an angular resolution $\sim 2\arcsec$, and a pixel scale of 1\farcs2.  GLIMPSE images have undergone zodiacal light subtraction based on a zodiacal emission model (see the data product manual\footnotemark[\value{footnote}]), so signal remaining in the mosaics is Galactic in nature.  There is, however, a significant effect due to scattering of light within the IRAC camera that causes the surface brightness of extended emission to appear brighter than it actually is\footnote{See \S4.11 of \texttt{http://irsa.ipac.caltech.edu/}\\\texttt{data/SPITZER/docs/irac/iracinstrumenthandbook/}} \citep{Reach:2005}.  The method used in this study to correct for scattered light is described in \S\ref{sec:ubc_proc}, and a derivation of the correction factors required for quantities measured from the publicly-available GLIMPSE mosaics is given in Appendix \ref{app:scatter}.


\section{Distance Probability Density Functions}\label{sec:dpdf}

\subsection{Approach and Utility}\label{dpdf:approach}

We introduce an automated distance determination technique for molecular cloud clumps that allows for the joint application of many individual distance estimation methods.  Bayes' Theorem provides a framework for creating distance probability density functions (DPDFs) for dust-continuum-identified molecular cloud clumps that encode the confidence in source distances.  Kinematic distances derived from \vlsr\ and a Galactic rotation curve constitute the likelihood functions in the Bayesian context.  Because these likelihoods are subject to the KDA, prior DPDFs based on ancillary data must be applied to constrain the distance estimates.  The posterior DPDF is simply the product of the likelihood with the priors, suitably normalized.  Relative amplitudes of the posterior DPDF at each distance along the line of sight (\dsun) correspond to the probability of the source being at that distance.

Within this framework, any number of prior DPDFs may be applied to constrain the distances to molecular cloud clumps.  This paper describes two such priors.  The first, applicable to all molecular cloud clumps, is based on the Galactic distribution of molecular hydrogen.  Because the scale height of the molecular disk is small, this prior favors the near kinematic distances for objects at high Galactic latitudes.  The second prior involves the use of EMAFs.  Not all molecular cloud clumps are visible as absorption features, however, so this prior (described in detail in \S \ref{sec:morph}) applies only to a subset of objects.  To expand the collection of molecular cloud clumps with well-constrained DPDFs, additional techniques (\eg HISA, NIREX, etc.) would need to be applied.

Not only do DPDFs provide a structure for applying multiple techniques for distance discrimination, they also encode the distance uncertainty and level of confidence in the KDA resolution.  When used to derive the mass or other property of a molecular cloud clump, DPDFs provide a means for determining the associated uncertainty.  The DPDFs derived in this work are computed out to a heliocentric distance of 20~kpc in 20-pc intervals.  To facilitate the use of integrated probabilities, DPDFs are normalized to unit total probability such that $\int_0^\infty \mathrm{DPDF}~\mathrm{d}(d_{_\sun}) = 1$.

\subsection{Extracting a Distance from the DPDF}\label{dpdf:dist}

The proper use of DPDFs for calculating derived quantities is to build a distribution by randomly sampling distances from the DPDFs in a Monte Carlo fashion, preserving all information about distance placement and uncertainty.  There are applications, however, that benefit from or require a single distance estimate with uncertainty (such as distance comparisons with other studies).  There are two primary distance estimates that may be derived from a DPDF.  The maximum-likelihood distance (\dml) is the distance which maximizes the DPDF.  This represents the single best-guess at the distance for cases where a large fraction of the total probability lies within a single peak.  The associated uncertainty may be defined as the confidence region around \dml\ that encloses at least 68.3\% of the integrated DPDF, and whose limits occur at equal relative probability.  This so-called isoprobability confidence region is generally asymmetric, and may represent lopsided error bars several kiloparsecs in size if both kinematic distance peaks are required to enclose sufficient probability.  The full width of this uncertainty (FW$_{68}$), therefore, provides a direct measure of how well constrained a distance estimate is.  Error bars produced in this way should not be considered Gaussian, as the 95.5\% and 99.7\% isoprobability confidence regions may be similar in size to the 68.3\% error bars, or be radically different.

An alternative single-value distance estimate is the weighted average distance (\dbar), the first moment of the distribution, 
\beqn\label{eqn:dbar}
\bar{d} = \int_0^\infty d_{_\sun}\ \mathrm{DPDF}\ \mathrm{d}(d_{_\sun})\ .
\eeqn
If the DPDF is well-constrained to a single peak, \dml\ and \dbar\ will be nearly equivalent.  In cases where the KDA resolution is not well-constrained, however, these distance estimates may be substantially different and \dbar\ is not a good estimator of the distance.  The uncertainty associated with \dbar\ may be computed from the second moment of the DPDF as
\beqn\label{eqn:sig_dbar}
\sigma_{_{\bar{d}}} = \left( \int_0^\infty d_{_\sun}^{\ 2}\ \mathrm{DPDF}\ \mathrm{d}(d_{_\sun})\ - \bar{d}\ ^2\right)^{1/2}\ .
\eeqn
The $\sigma_{_{\bar{d}}}$ represent the variance of the DPDF, and only approximate Gaussian confidence intervals for single-peaked DPDFs.  Ultimately, the choice of a single-value distance estimate will depend on the specifics of the application; various cases are discussed in \S\ref{disc:dist_est}.

\subsection{Using DPDFs to Estimate Physical Parameters}\label{dpdf:physical}

While distances to objects are often interesting in isolation, their primary use is to convert observational quantities into physical properties of the object.  DPDFs offer a simple way to propagate the uncertainties in distance through these calculations.  For example, the maximum-likelihood mass of a molecular cloud clump can be estimated as
\beqn\label{eqn:mass_single}
M_{_\mathrm{ML}} = \alpha~S_{_{1.1}}~d_{_\mathrm{ML}}^{~2}~,
\eeqn
where $S_{_{1.1}}$ is the $\lambda = 1.1$~mm flux density, and $\alpha$ contains the dust physics and temperature.  Adoption of a DPDF representation allows marginalization over distance to obtain the expectation value of the mass:
\beqn\label{eqn:mass_dpdf}
\langle M \rangle = \int_{0}^{\infty}  \alpha~S_{_{1.1}}~d_{_\sun}^{\ 2}~\mathrm{DPDF}~\mathrm{d}(d_{_\sun})~.
\eeqn
Practically, this integration can be accomplished by Monte Carlo methods, drawing a large number of distance samples from the DPDF and evaluating the average mass.  Uncertainties in the expectation value can be determined using methods paralleling those used for distance above.

Bimodal DPDFs again lead to complications, as the expectation value will commonly be found at a value with low probability.  A maximum likelihood distance can be adopted to avoid this aesthetic feature, but marginalization over the distance remains the most rigorous approach.  Ideally, additional prior DPDFs should be applied in order to minimize bimodality.

\subsection{Kinematic Distance DPDFs}\label{dpdf:kdist}

Kinematic distances form the foundation for the Bayesian approach to distance estimation, computed from the intersection of the Galactic rotation curve projected along the line of sight, $v($\dsun$)$, with the observed molecular line \vlsr.  Transformation of velocity uncertainties onto the distance axis is facilitated by the use of two-dimensional probability density functions, $P$(\vlsr,~\dsun).  

The rotation curve function, $P_\mathrm{rotc}$(\vlsr,~\dsun), is constructed as
\beqn\label{eqn:protc}
P_\mathrm{rotc}(v_{_\mathrm{LSR}},d_{_\sun}) = \exp \left(- \frac{ \left[ v_{_\mathrm{LSR}} - v(d_{_\sun})\right]^2}{2 \sigma_\mathrm{vir}^2}  \right )\ ,
\eeqn
where the uncertainty $\sigma_\mathrm{vir}$ is the magnitude of expected virial motions within regions of massive-star formation, accounting for peculiar motions of individual molecular cloud clumps \citep[=~7~\kms;][]{Reid:2009}\footnote{This is the expected virial velocity, per coordinate, for an individual object (\ie molecular cloud clump) within a high-mass star-forming region of mass $\sim$3$\times10^4~M_\sun$ and radius $\sim$1~pc \citep{Reid:2009}.}.  The function is Gaussian in \vlsr, and is centered along $v($\dsun$)$; if integrated over \vlsr, a uniform DPDF is obtained.  The probability density function from spectral line information ($P_\mathrm{spec}$) is a Gaussian centered at the measured $v_\mathrm{line}$, with observed linewidth $\sigma_\mathrm{line}^2$, independent of \dsun.  As with $P_\mathrm{rotc}$, this function yields a uniform DPDF when integrated over \vlsr.  Since $P_\mathrm{rotc}$ does vary as a function of \dsun, localized peaks in the (\vlsr,~\dsun) plane result when it is multiplied by $P_\mathrm{spec}$.  The desired one-dimensional DPDF$_\mathrm{kin}$ is obtained by subsequent integration over \vlsr.

DPDF$_\mathrm{kin}$ is double-peaked and symmetric about the tangent distance for objects with $R_\mathrm{gal} < R_0$, and single-peaked otherwise.  The $v($\dsun$)$ were computed using the flat rotation curve of \citet{Reid:2009}.  \citet{Schonrich:2010} subsequently derived newer estimates of the Solar peculiar motion, affecting rotation curve fits to the maser parallax data of \citeauthor{Reid:2009}  The updated values used here are $R_0 = 8.51$~kpc, and $\Theta_0 = 244$~\kms\ (M. Reid 2011, private communication).  The new solar motion values also had the effect of decreasing the magnitude of the apparent Galactic counter-rotation of high-mass star forming regions, an effect likely arising from molecular gas interacting with the spiral potential, from 15~\kms\ to 6~\kms.

Kinematic distances are sensitive to the slope of $v($\dsun$)$, itself a function of Galactic longitude.  For lines of sight along $b \approx0\degr$ within $\sim 10\degr$ of the Galactic longitude cardinal directions, $v($\dsun$)$ is either very flat or sharply peaked; small departures from circular motion therefore translate into large deviations in derived kinematic distances.  Furthermore, since $v($\dsun$)$ is derived assuming circular orbits about the Galactic center, radial streaming motions of the gas are not accounted for, meaning that DPDF-derived distance estimates carry the basic limitations of any kinematic distance determination.  To minimize the effects of non-circular motion, regions known to have significant streaming must be excluded from consideration.  In particular, the presence of the long Galactic bar at $R_\mathrm{gal} \lesssim 3$~kpc \citep{Fux:1999,RodriguezFernandez:2008} and its associated radial streaming motions restrict the use of kinematic distance measurements to locations outside this radius.  In the Galactic longitude-velocity ($\ell-v$) diagram, these restrictions amount to excluding much of $|\ell| \lesssim 20\degr$.  Features at low longitude known to be outside the Galactic bar \citep[such as the Scutum-Centarus arm, also labeled as the ``Molecular Ring'';][their Fig. 3]{Dame:2001}, may be considered to have roughly circular orbits, and are included in this study.

\subsection{Prior DPDFs for Kinematic Distance Discrimination}\label{dpdf:priors}

Prior DPDFs are required to discriminate between the kinematic probability peaks for objects within the solar circle.  DPDF$_\mathrm{kin}$ is symmetric about the tangent point, so prior DPDFs based on ancillary Galactic plane data must be asymmetric to provide useful distance constraints.  

The Galactic distribution of molecular gas serves as an envelope inside which molecular cloud clumps may form.  The prior DPDF$_{\mathrm{H}_2}$ is defined to be proportional to the volume density from the molecular hydrogen model of \citet{Wolfire:2003} along a line of sight.  This model consists of a Molecular Ring component with a decaying exponential toward the outer Galaxy; the vertical distribution is Gaussian with a half-width at half maximum of 60~pc \citep{Bronfman:1988}, flaring outside the Solar Circle.  While this distribution is symmetric about \dtan\ along the Galactic midplane, the narrow vertical extent of the molecular layer sets a strong prior on higher-latitude objects.  The relative amount of \htwo\ beyond the tangent point for lines of sight at $|b| \gtrsim 0\fdg3$ is small, generating the needed asymmetric function for molecular cloud clumps at larger Galactic latitude.

The prior DPDF based on EMAFs was computed from a pixel-by-pixel morphological matching between millimeter dust-continuum emission and mid-infrared dust absorption features.  The derivation of DPDF$_\mathrm{emaf}$ is described in detail in the next section.


\section{Infrared-Millimeter Morphological Matching}\label{sec:morph}

Morphological matching is based on the comparison between synthetic 8-\micron\ images computed from millimeter flux density measurements and GLIMPSE 8-\micron\ maps processed to match the angular resolution of the BGPS.  This section describes the creation of both the synthetic and processed 8-\micron\ images, as well as the mechanics of computing DPDF$_\mathrm{emaf}$.

\subsection{Creation of Synthetic 8-\micron\ Images}\label{morph:formal}

\subsubsection{Radiative Transfer Assumptions}

Creation of  synthetic 8-\micron\ images explicitly assumes that the dust seen in emission in the BGPS is the same dust that extincts mid-infrared light.  When converted into a mid-infrared optical depth, BGPS observations represent dark clouds which may be placed at different heliocentric distances within a model of diffuse Galactic 8-\micron\ emission.  A series of synthetic images generated in this manner were compared with mid-infrared observations to compute the DPDF$_\mathrm{emaf}$.

We assumed a simple radiative transfer model to describe the observed mid-infrared intensity absorbed by a cold molecular cloud clump immersed in a sea of diffuse emission (assuming that the absorbing cloud has no emission).  The intensity observed within an EMAF ($I_\mathrm{emaf}$) is
\beqn\label{eqn:rad_xfer}
I_\mathrm{emaf} = I_\mathrm{back}\ e^{-\tau_{_8}} + I_\mathrm{fore}\ ,
\eeqn
where $I_\mathrm{back}$ and $I_\mathrm{fore}$ are the background (from the cloud to large heliocentric distance) and foreground (between the observer and the cloud) intensities, respectively, and $\tau_{_8}$ is the mid-infrared optical depth of the cloud.  The total intensity along a line-of-sight in the absence of absorption is $I_{_\mathrm{MIR}} = I_\mathrm{back} + I_\mathrm{fore}$.  Defining the fraction of the total intensity that lies in front of the cloud as $f_\mathrm{fore} = I_\mathrm{fore} / I_{_\mathrm{MIR}}$ allows Equation (\ref{eqn:rad_xfer}) to be written as
\beqn\label{eqn:imin_model}
I_\mathrm{emaf} = \left[(1 - f_\mathrm{fore})\ e^{-\tau_{_8}} + f_\mathrm{fore}\right]\ I_{_\mathrm{MIR}}\ .
\eeqn
This parameterization frames the observed EMAF intensity in terms of decrements below the un-extincted intensity in the vicinity, and provides the basis for creating synthetic 8-\micron\ images.  It follows quickly from Equation~(\ref{eqn:imin_model}) that clouds optically thick in the mid-infrared ($\tau_{_8}\sim1$) will still have a 10\% difference between $I_\mathrm{emaf}$ and $I_{_\mathrm{MIR}}$ (\ie easily detectable) for \ffore\ as large as 0.85.  Calculation of $\tau_{_8}$ and \ffore\ are described below, and the estimation of $I_{_\mathrm{MIR}}$ from GLIMPSE data is discussed in \S \ref{sec:ubc_proc}.

\subsubsection{8-\micron\ Optical Depth from the Millimeter Flux Density}\label{morph:tau_8}

The mid-infrared optical depth of an EMAF cannot be measured directly from the GLIMPSE mosaics without significant assumptions, but it may be estimated from millimeter data.  Thermal dust emission is optically thin at millimeter wavelengths, so the observed BGPS flux density ($S_{_{1.1}}$) may be written as
\beqn\label{eqn:s_bgps}
S_{_{1.1}} = B_{_{1.1}}(T_d)\ \tau_{_{1.1}}\ \Omega_{_\mathrm{BGPS}}\ ,
\eeqn
where $B_{_{1.1}}(T_d)$ is the Planck function evaluated at $\lambda = 1.1$~mm and dust temperature $T_d$, and $\Omega_{_\mathrm{BGPS}} = 2.9 \times 10^{-8}$~sr is the solid angle of the BGPS beam.  The millimeter optical depth ($\tau_{_{1.1}}$) was computed assuming the dust opacity ($\kappa_{_{1.1}}$) for grains with thin ice mantles, coagulating at 10$^6$~cm$^{-3}$ for 10$^5$~years \citep[][Table 1, Column 5; called OH5 dust]{Ossenkopf:1994}.  Interpolation of OH5 dust opacities to the central frequency of the BGPS bandpass yields $\kappa_{_{1.1}} = 1.14$~cm$^2$~g$^{-1}$ of dust (A11).  A molecular cloud clump with $\tau_{_{1.1}} = 10^{-3}$, which corresponds to ($S_{_{1.1}} \approx 0.9$~Jy), has a beam-averaged molecular hydrogen column density $\approx 2\times10^{22}$~cm$^{-2}$.

The 8-\micron\ optical depth is related to $\tau_{_{1.1}}$ by the ratio of the dust opacities in the two bandpasses, $R_\kappa = \kappa_{8} / \kappa_{_{1.1}}$.  We calculated the mid-infrared dust opacity by assuming a dust emission spectrum including PAH molecules \citep{Draine:2007}, finding the average attenuated intensity across IRAC Band~4, and extracting a band-averaged opacity $\kappa_{8}=825$~cm$^2$~g$^{-1}$ of dust (see Appendix~\ref{app:dust_opacity}).  At the 33\arcsec\ resolution of the BGPS, the beam-averaged 8-\micron\ optical depth is therefore
\begin{eqnarray}\label{eqn:tau_8}
\tau_{_8} &=& \frac{R_\kappa}{B_{_{1.1}}(T_d) \ \Omega_{_{\mathrm{BGPS}}}}\ S_{_{1.1}} = \Upsilon(T_d) \ S_{_{1.1}} \nonumber \\
&=& 0.778\ \left( \frac{e^{13.0\mathrm{K}/ T_d}-1}{e^{13.0\mathrm{K}/20.0\mathrm{K}}-1} \right) \left(\frac{S_{_{1.1}}}{\mathrm{1\ Jy}}  \right)\ .
\end{eqnarray}
The function $\Upsilon(T_d)$ has units of inverse flux density, and is normalized to 20~K in Equation~(\ref{eqn:tau_8}).  Because $\tau_{_8}$ is a function of $R_\kappa$ (\ie both millimeter-wave emission and mid-infrared absorption depend only on the dust), the dust-to-gas ratio is not relevant to the distance estimation method.  Owing to the nearly three orders of magnitude difference in dust opacity between the millimeter and mid-infrared, a value of $\tau_{_8} = 0.1$ corresponds to a column of only $N($\htwo$) \approx 3 \times 10^{21}$~cm$^{-2}$, assuming $A_{[8\mu]}/A_V \approx 0.05$ \citep{Indebetouw:2005,RomanZuniga:2007}.  Therefore, molecular cloud clumps with column densities $\gtrsim 10^{22}$~cm$^{-2}$ will be mostly opaque at $\lambda = 8$~\micron.

Using Equation (\ref{eqn:tau_8}) to obtain an 8-\micron\ optical depth requires a dust temperature ($T_d$).  Since we are ignorant of $T_d$ within each molecular cloud clump used in this study, we assumed that all sources are at the same temperature.  \citet{Battersby:2011} showed that mid-infrared-dark molecular cloud clumps generally span the temperature range 15~K $\lesssim T_d \lesssim$ 25~K.  Therefore, $T_d = 20$~K is a reasonable representation for BGPS sources as a group.  Variation of the assumed $T_d$ affects the KDA resolutions for some sources, and is discussed briefly in \S\ref{disc:discrim}.  With molecular cloud clump dust temperatures derived from {\it Herschel} Hi-GAL data, more precise DPDFs for individual objects may be derived using the present methodology.

\subsubsection{8-\micron\ Foreground Fraction from a Galactic Emission Model}\label{morph:model}

Absorption features seen at $\lambda = 8$~\micron\ are assumed to be the result of dense clouds immersed in a smooth emission distribution, punctuated by regions undergoing active star formation.  While small-scale structures are difficult to model, the broader diffuse emission is a more tractable problem.  Creation of synthetic 8-\micron\ images via Equation~(\ref{eqn:imin_model}) requires a three-dimensional model for the Galactic 8-\micron\ emission distribution.

The recent numerical Galactic stellar and dust emission model of \citet[][hereafter R12]{Robitaille:2012}, computed using the Monte-Carlo three-dimensional radiative transfer code {\sc Hyperion}\footnote{\texttt{http://www.hyperion-rt.org}} \citep{Robitaille:2011}, offers a self-consistent estimate of diffuse Galactic emission that is well-matched to observed quantities.  We used the final model presented in R12, whose parameters were chosen to fit the Galactic latitude and longitude intensity distributions from seven bandpasses in the mid- to far-infrared.  This model features two major and two minor spiral arms with Gaussian radial profiles, a lack of dust in the inner few kiloparsecs of the Galactic disk (dust hole; correlated with the dearth of molecular gas in this region), and a modified PAH abundance relative to the favored model from \citet{Draine:2007}.  An analysis of the contributions from various stellar populations and dust grain sizes to the total intensity in each bandpass indicates that some 96\% of the emission detected in IRAC Band~4 images comes from PAH molecules (R12).

\begin{deluxetable*}{llcc}
  \tablecolumns{4}
  \tablewidth{0pt}
  \tabletypesize{\small}
  \tablecaption{Comparison of {\sc Hyperion} Model Parameters\label{table:hyperion}}
  \tablehead{
    \colhead{Category} & \colhead{Parameter} & \colhead{R12} & \colhead{This Work}
  }
  \startdata
  Grid\tablenotemark{a} & $N_R$       & 200 & 200 \\
       & $N_\phi$  & 100 & 200 \\
       & $N_z$       & 50  & 44  \\
       & $|z|_\mathrm{max}$ (pc) & 3000 & 1000 \\
	\hline
  Wavelength\tablenotemark{b} & $N$ bins   & 160 & 22 \\
             & Range (\micron)  & $3 \leq \lambda \leq 140$ & $6 \leq \lambda \leq 10$ \\
	\hline
  Image\tablenotemark{c} & Observer $R_\mathrm{gal}$ (kpc) & 8.5 & 8.5 \\
        & Observer $z$ (pc) & $+15$ & $+25$ \\
        & Longitude Range (\degr)  & $65 \geq \ell \geq -65$ & $65 \geq \ell \geq -65$
  \enddata
  \tablenotetext{a}{$N$ = number of grid cells in this dimension}
  \tablenotetext{b}{Wavelengths at which the model images were computed later convolved with instrument bandpasses to create simulated observations.}
  \tablenotetext{c}{Parameters related to observer within the grid.}
\end{deluxetable*}

\begin{figure}[t]
  \centering
  \includegraphics[width=3.1in]{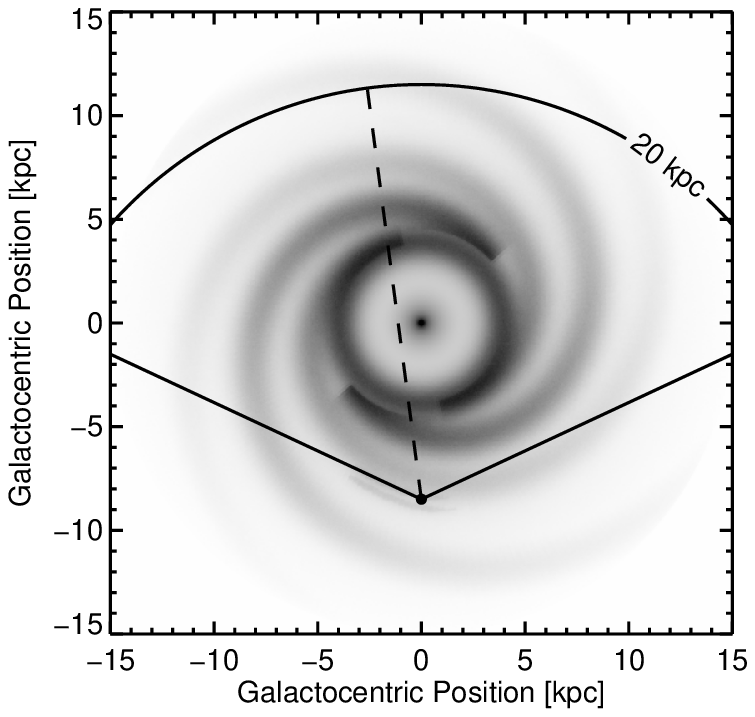}
  \caption{Galactic mid-infrared emission model computed with {\sc Hyperion} \citep{Robitaille:2012} viewed from the North Galactic Pole.  The model, viewed through the IRAC 8.0-\micron\ bandpass, is shown on an inverted square-root intensity scale.  The Sun is located at $(x,y) = (0,-8.5\ \mathrm{kpc})$, and solid diagonal lines represent the limits of the GLIMPSE survey ($|\ell| = 65\degr$).  The dashed line marks the low-latitude ($\ell=7\fdg5$) limit of this study.  DPDFs were computed out to \dsun~=~20~kpc (curved contour).}
  \label{fig:galmodel}
\end{figure}

A three-dimensional ($\ell,b,$\dsun) data cube of 8-\micron\ emission was generated from the radiative transfer code using model grid and image parameters slightly modified from those used by \citeauthor{Robitaille:2012}  Table~\ref{table:hyperion} lists the comparison of {\sc Hyperion} input parameters between R12 and the present study.  Primary differences include an increase in azimuthal resolution of the cylindrical grid, a restriction of the vertical extent of the grid to $|z| \leq 1$~kpc (to match the region of the Galactic plane probed by the latitude range $|b| \leq 1\degr$), and limiting the wavelength range used in computing output images.  The model was computed within a box 30~kpc on a side, containing the entire modeled stellar disk (R12); a face-on view of the model Milky Way as seen from the north Galactic pole is shown in Figure~\ref{fig:galmodel}.  Lines of sight out to 20~kpc (the distance used for DPDF generation) lie entirely within the simulation box for $|\ell|\leq48\degr$.  Beyond this longitude, however, the edge of the box retreats to only \dsun~$\approx 16.5$~kpc by $|\ell|=65\degr$.

A series of ($\ell,b$) images of the Galactic plane containing only emission between the observer and some distance $d_i$ were computed using a useful {\sc Hyperion} feature, using a resolution of 3\arcmin\ in latitude, and 15\arcmin\ in longitude for computational reasons.  Images were computed for each of 22 wavelength bins logarithmically spaced from 6 to 10~\micron\ (closely matching the wavelength bins used by R12 for this part of the spectrum), and were then convolved with the IRAC Band~4 transmission curve to yield a single simulated \spitzer\ image (see \citealp{Robitaille:2007} for complete details).  The three-dimensional image was constructed by stepping $d_i$ outward in 100-pc intervals and depicts the cumulative 8-\micron\ emission out to each $d_i$.  The resulting cube comprises 200 steps, with the final image slice including all model emission to the edge of the box, equivalent to the collapsed profiles presented in R12.

{\sc Hyperion} cannot treat sources individually, but rather uses ``diffuse'' sources of emission in each grid cell.  These diffuse sources are generated from the probability of emission from various populations of stars and similar objects (\eg planetary nebulae, \HII\ regions), assigned a spectrum corresponding to the appropriate spectral class, and given a total luminosity based on the number of ``real'' sources the cell represents.  While most of the emitting populations have a smooth spatial distribution, relatively rare sources with concentrated emission at $\lambda = 8$~\micron\ (such as \HII\ regions) are sprinkled throughout the box according to the underlying stellar distribution model.  Very nearby objects (\dsun~$\leq 0.5$~kpc) appear quite bright, and cause ``hot-pixel'' effects in the computed images of the Galactic plane.  These objects blend into the background for images computed from large Galactocentric position (\eg Fig.~\ref{fig:galmodel}), or are averaged out in collapsed longitude or latitude distributions (R12).  To ameliorate the effect of these objects in the computed $(\ell,b)$ images, we ran seven realizations of the model, each with a different random-number seed, then median-combined the realizations of each $d_i$ slice.  Since the underlying distribution of sources is fixed, nearby bright sources often appear in the same pixel in the output images; the number of realizations was chosen to be large enough such that median combining the realizations removes most of these outliers.  To eliminate any remaining outliers and reduce noise, the combined ($\ell,b$) images were median smoothed with a 3 pixel $\times$ 3 pixel box.

\begin{figure}[t]
  \centering
  \includegraphics[width=3.1in]{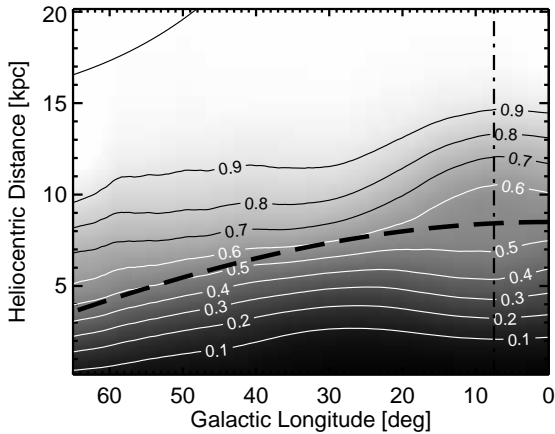}
  \caption{Foreground fraction of Galactic 8-\micron\ emission in the northern Galactic plane derived from the {\sc Hyperion} model as a function of ($\ell,$\dsun) along $b = 0\degr$.  Grayscale and contours represent \ffore, with the unlabeled 1.0 contour marking the edge of the box in Fig.~\ref{fig:galmodel} for $\ell \gtrsim 48\degr$.  The thick black dashed line follows the tangent distance as a function of Galactic longitude, and the vertical dot-dashed line marks the $\ell = 7\fdg5$ lower limit of this study.}
  \label{fig:3d_ffore}
\end{figure}

The foreground fraction was computed from the intensity cubes by dividing each ($\ell,b$) image slice by the final slice. The final {\sc fits} data cubes of 8-\micron\ intensity and \ffore\ for both the northern and southern Galactic plane ($|\ell| \leq 65\degr$) are publicly available with the BGPS archive.  To illustrate the Galactic features present in the modeled cube, \ffore($\ell$,\dsun) for the Northern plane along $b=0\degr$ is shown in Figure~\ref{fig:3d_ffore}, with contours and grayscale representing its value from 0 to 1.  Since PAH molecules contribute the bulk of the model emission, the dust hole towards low longitude is visible as a flattening of \ffore(\dsun).  The Molecular Ring / Scutum tangent at $\ell \approx 30\degr$ appears where \ffore\ grows quickly as a function of distance.  The limited distance range caused by the model box size is represented by the 1.0 contour for $\ell\gtrsim48\degr$.  The tangent distance as a function of longitude (black dashed line) spans the range $0.45\lesssim f_\mathrm{fore} \lesssim 0.6$, implying that clouds that are optically thick in the mid-infrared should be visible beyond \dtan.

\subsubsection{Computing the Synthetic Images}

Synthetic images ($I_\mathrm{emaf}$) for a given BGPS object are computed using Equation~(\ref{eqn:imin_model}).  The optical depth is modeled as a two-dimensional image, constructed by applying Equation~(\ref{eqn:tau_8}) to the BGPS map data.  The estimate of the total mid-infrared emission ($I_{_\mathrm{MIR}}$) is also a two-dimensional image, and its creation is discussed below.  Because of the coarse resolution of the \ffore\ model, we simply extracted the one-dimensional \ffore(\dsun) at the ($\ell,b$) of the BGPS object.  The combination of these elements yields a cube of synthetic data to be compared with the processed GLIMPSE images.

\subsection{Processing of GLIMPSE 8-\micron\ Images}\label{sec:ubc_proc}

\begin{figure*}[t]
        \centering
        \includegraphics[width=6.5in]{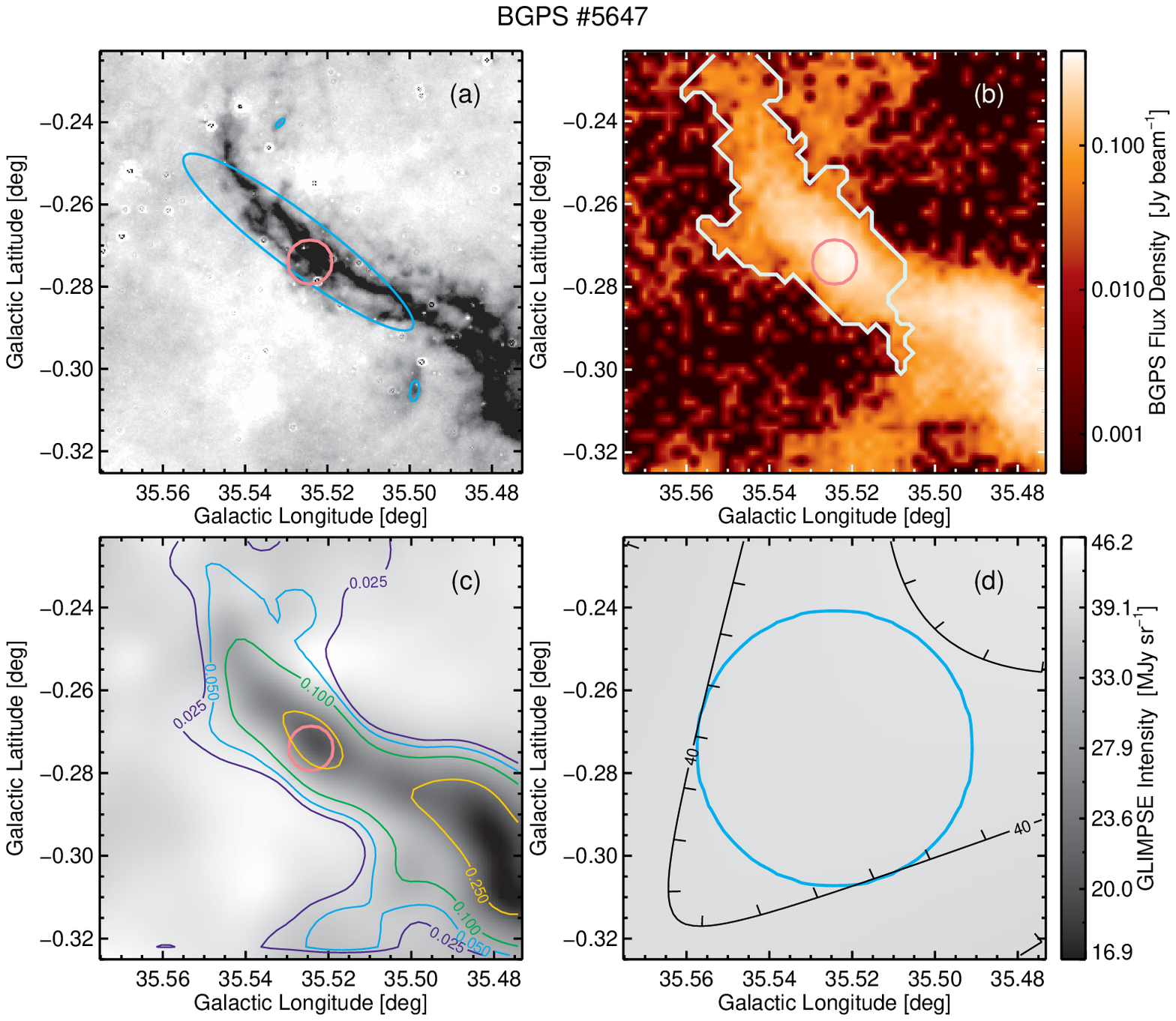}
        \caption{BGPS and processed GLIMPSE data for example object G035.524-00.274.  All panels are $6\arcmin\times6\arcmin$ postage-stamp images (see text), and the pink circle identifies the 40\arcsec\ top-hat BGPS equivalent aperture, centered on the location of peak flux density.  ({\it a}) Cutout of the star-subtracted GLIMPSE image at native resolution.  Cyan ellipses mark IRDCs identified in the \citet{Peretto:2009} catalog; note that the dark cloud associated with BGPS source G035.478-00.298 (lower-right corner) is not included in that catalog.  ({\it b}) BGPS map data with Bolocat source boundary (light cyan).  ({\it c})  Star-subtracted GLIMPSE cutout smoothed to 33\arcsec\ and resampled to 7\farcs2 pixels to match the BGPS maps.  Color contours represent logarithmic flux density levels from BGPS (Jy~beam$^{-1}$).  ({\it d}) Estimate of the total mid-infrared intensity, $I_{_\mathrm{MIR}}$, as a quadratic surface fitted to background pixels as described in the text.  Contours are drawn to show the variation in $I_{_\mathrm{MIR}}$ over the postage stamp (MJy~sr$^{-1}$; ticks point to higher values), and the cyan aperture marks the region used to estimate $\langle I_{_\mathrm{MIR}} \rangle$ for this source (see \S\ref{res:sources}).  The grayscale colorbar represents the common (logarithmic) intensity scale for all three GLIMPSE panels.  (A color version of this figure is available in the online journal.)}
        \label{fig:glimpse_ubc}
\end{figure*}

Mid-infrared properties of dust-continuum-identified molecular cloud clumps were derived from the \spitzer/GLIMPSE mosaics.  Further processing of these images was required to estimate the total mid-infrared intensity ($I_{_\mathrm{MIR}}$) in the vicinity of an EMAF, and to produce a smoothed, star-subtracted map, containing features and angular scales comparable to (sub-)millimeter data.  The example source G035.524-00.274 (BGPS \#5647) is used to illustrate the processing products in Figure~\ref{fig:glimpse_ubc}.  The first step was to remove individual stars because they contaminate estimates of broader diffuse emission and  (sub-)millimeter observations are not sensitive to them.  Star locations were identified by searching for bright, unresolved objects in the 3.6-\micron\ mosaics using \texttt{DAOFIND} \citep{daophot,astrolib} with a threshold of 20 MJy~sr$^{-1}$.  A Gaussian was fit to the 8-\micron\ image at the location of each identified star, then subtracted.  This method of star subtraction was deemed optimal because PSF variations across the survey mosaics meant that PSF-based approaches could not be applied.  Star subtraction in this manner did, however, leave clear low-level residuals (Fig.~\ref{fig:glimpse_ubc}{\it a}).  Since later processing smooths the resulting images to the BGPS resolution, residuals are largely unimportant.  However, to ensure that poor star subtraction or other effects did not effect distance estimation, by-eye evaluation of each potential EMAF for contamination was performed.

For further processing of the GLIMPSE data, $6\arcmin \times 6\arcmin$ postage-stamp images were extracted from the star-subtracted 8-\micron\ mosaics for each Bolocat source.  These postage stamps, centered on the location of peak millimeter flux density, limit consideration of mid-infrared variations to the immediate vicinity of a molecular cloud clump in addition to providing computational expediency.  The first postage-stamp image created for a given BGPS object is a version of the star-subtracted GLIMPSE mosaic, re-pixelated and aligned to the 7\farcs2 scale of the BGPS images.  This image was used to ensure that locally bright emission did not interfere with derived mid-infrared intensities, and to determine the likely intensity range containing $I_{_\mathrm{MIR}}$ around the object.  A pixel intensity histogram of the image was constructed with 1 MJy~sr$^{-1}$-wide bins, and the background was defined as intensities within its full-width at half maximum.  Pixels within an $8\arcmin\times8\arcmin$ section of the native-resolution star-subtracted GLIMPSE mosaic having intensities in the defined range were used to fit a quadratic surface using a linear, least-squares optimization.  This surface, repixelated and scaled as above, comprises the postage-stamp estimate of $I_{_\mathrm{MIR}}$ (Fig.~\ref{fig:glimpse_ubc}{\it d}).  This estimate of background pixels ignored high pixel values from star residuals and low pixel values from EMAFs.

The IRAC camera on \spitzer\ suffers from internal scattering of light which affects instrument calibration \citep{Reach:2005}.  Point-source photometry is unaffected by the scattering due to the calibration technique employed, but extended emission (such as the Galactic plane) will appear brighter due to scattering into each pixel.  Correcting for this effect should be done on a frame-by-frame basis, but was not accounted for in the GLIMPSE pipeline (S. Carey 2010, private communication).  To approximately correct for the scattering, an estimate of the scattered light was subtracted from the postage-stamp images for each BGPS source.  The postage-stamp size was chosen to be near the $5\farcm2 \times 5\farcm2$ FOV of IRAC, and the $I_{_\mathrm{MIR}}$ fit serves as the estimate of the light available to be scattered within a single IRAC frame.  This estimate is only approximate, as the $I_{_\mathrm{MIR}}$ fit explicitly excludes very bright and very dim emission within a frame; for frames with regions of bright emission, the derived correction factor will be a lower limit, and vice verse for frames containing extensive dark clouds.    The infinite-aperture intensity correction for IRAC Band~4 is 0.737 \citep{Reach:2005}, meaning that $\xi = 0.263$ is the scattered light fraction.  Assuming that $I_{_\mathrm{MIR}}$ represents the total incident light, we subtracted $\xi \times \mathrm{median}(I_{_\mathrm{MIR}})$ from each postage-stamp image to remove scattered light.

Reduction of the GLIMPSE angular resolution was necessary for direct comparison with the synthetic images created using Equation (\ref{eqn:imin_model}).  Since bright emission in the vicinity is scattered into an EMAF, removal of the scattered light must be done prior to smoothing.  The scattering-corrected extracted postage stamps were smoothed with a FWHM = 33\arcsec\ Gaussian kernel, then re-pixelated and aligned to match the BGPS images (Fig.~\ref{fig:glimpse_ubc}{\it c}).

\subsection{Morphological Matching}\label{morph:morph}

\begin{figure*}[t]
        \centering
        \includegraphics[width=6.5in]{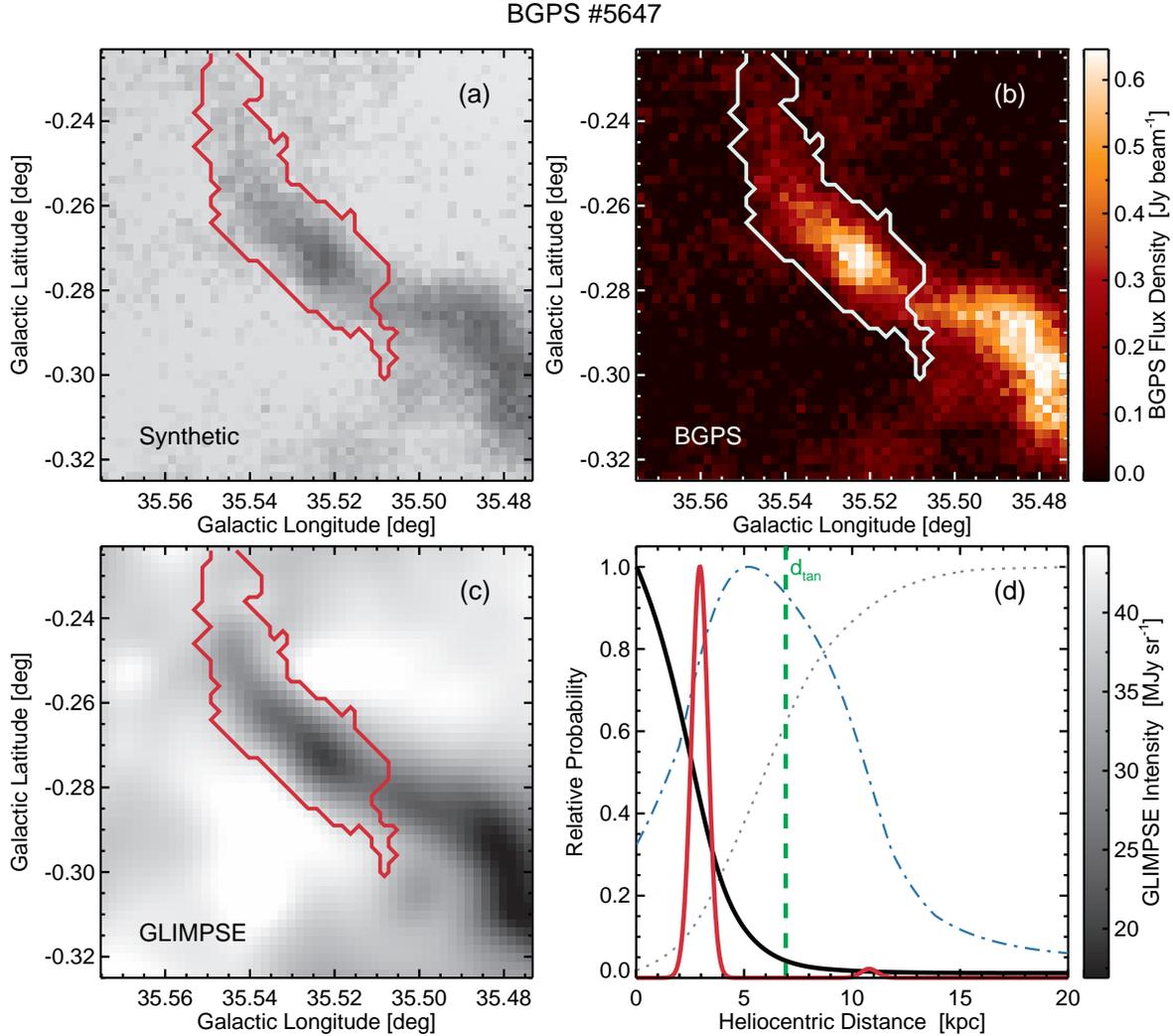}
        \caption{Morphological matching for example object G035.524-00.274.  The millimeter source to the lower-right of the marked contour is a separate Bolocat object.  ({\it a}) Synthetic 8-\micron\ image calculated via Equation~(\ref{eqn:imin_model}).  ({\it b}) BGPS postage-stamp image, showing the restricted region used for the morphological matching (see text).  ({\it c}) Smoothed GLIMPSE map against which the synthetic images are compared.  ({\it d}) Prior DPDFs from the morphological matching (black) and molecular gas distribution (blue dot-dashed), and the posterior DPDF (red), including the kinematic distance likelihood (see text).  The gray dotted line represents \ffore\ extracted from the model cube along ($\ell,b$), and the green dashed line marks the tangent distance.  (A color version of this figure is available in the online journal.)}
        \label{fig:morph}
\end{figure*}

Derivation of DPDF$_\mathrm{emaf}$ relies upon the comparison of the (sub-)millimeter emission and mid-infrared absorption of dust in cold molecular cloud clumps.  The series of synthetic images were matched against the smoothed GLIMPSE postage stamp images (Fig.~\ref{fig:glimpse_ubc}{\it c}).  For small \dsun, the synthetic cloud appears darkest, without foreground light filling in the absorption feature.   At larger heliocentric distance, \ffore\ grows, and the synthetic image converges upon $I_{_\mathrm{MIR}}$ (Fig.~\ref{fig:glimpse_ubc}{\it d}).

The Galactic 8-\micron\ emission model of \S\ref{morph:model} describes smooth, diffuse emission against which EMAFs are visible, but actual Galactic emission is more complex.  To match more closely the assumption of the model, the angular region over which the synthetic and observed sky are compared must be restricted.  As the observational definition of a single molecular cloud clump, we began with a source's Bolocat contour delineating the maximum extent of the comparison region \citep{Rosolowsky:2010}.  Since the synthetic image can never be brighter than $I_{_\mathrm{MIR}}$, the matching process is adversely affected by bright mid-infrared emission in the vicinity of a BGPS source.  To ameliorate this effect, pixels in the smoothed GLIMPSE postage stamp were excluded from the matching region if their value exceeded the corresponding value in the $I_{_\mathrm{MIR}}$ image.

An overview of the morphological matching process is presented in Figure~\ref{fig:morph} for the same object as in Figure~\ref{fig:glimpse_ubc}.  The synthetic 8-\micron\ image is shown in panel ({\it a}) for the distance which maximizes DPDF$_\mathrm{emaf}$ (see below).  Panels ({\it b}) and ({\it c}) are identical to Figure~\ref{fig:glimpse_ubc}, except that the source contour now marks the restricted matching region due to bright mid-infrared emission on the perimeter of the EMAF.  Panels ({\it a}) and ({\it c}) are shown on a common linear grayscale to illustrate the match between the observed extinction and that predicted from thermal dust emission.  The various DPDFs for this source are shown in panel ({\it d}), and are described below.

Quantification of the match as a function of distance was accomplished by constructing a $\chi^2$ statistic from a pixel-by-pixel comparison within the matching region.  The estimate of the error in each pixel was derived from Equation~(\ref{eqn:imin_model}) by propagating the uncertainty in the optical depth map as
\beqn
\sigma_\mathrm{syn}(\ell,b) = I_{_\mathrm{MIR}}(\ell,b)\ e^{-\tau_{_8}(\ell,b)}\ \Upsilon\ \sigma_{_\mathrm{S_{_{1.1}}}}\ ,
\eeqn
where Equation~(\ref{eqn:tau_8}) defines $\tau_{_8}$ and $\Upsilon$, and $\sigma_{_\mathrm{S_{_{1.1}}}}$ is the median absolute deviation of the BGPS postage-stamp image.
This estimate of the uncertainty places more weight on the portions of the image with larger BGPS flux density.  The statistic was computed for synthetic images at 100-pc intervals along the line of sight, yielding $\chi^2($\dsun$)$.  

A preliminary DPDF$_\mathrm{emaf}$ was computed using the formal probability of the $\Delta \chi^2$ statistic.  The number of degrees of freedom was taken as the integer number of BGPS beams in the matching region ($N_\mathrm{pixels}$~/ 23.8~pixels~beam$^{-1}$; A11) minus one, since only beam-scale structures are independent and distance is a fitted parameter.  For most sources, the DPDF$_\mathrm{emaf}$ has a broad peak (several kiloparsecs wide), and falls sharply where the $\Delta (\chi^2_\mathrm{red})$ exceeds unity.  Because of the sharp cutoffs, it tends to very strongly favor one kinematic distance peak over the other.  If the Galaxy truly consisted of dark molecular cloud clumps embedded within broad diffuse mid-infrared emission, this formulation of DPDF$_\mathrm{emaf}$ would be appropriate.  However, the Galaxy is punctuated with regions of stronger 8-\micron\ emission that violate the simple radiative transfer of Equation~(\ref{eqn:rad_xfer}), and the DPDF$_\mathrm{emaf}$ should contain a systematic uncertainty that allows non-vanishing probability at the non-favored kinematic distance peak.

Experimentation with alternative approaches that allow a systematic uncertainty led to the selection of DPDF$_\mathrm{emaf} \propto (\chi^2)^{-\beta}$, where $\beta$ is a positive scalar of order unity.  This class of DPDF$_\mathrm{emaf}$ have FWHM comparable to the formal probability, but greater width at low likelihood, and hence rarely goes to zero probability until far from the peak.  The parameter $\beta$ may be used to tune the width of the function, with larger values leading to narrower distributions.  Since the sharp cutoff of the  DPDF$_\mathrm{emaf}$, not the width of the peak, is what appears problematic in light of complex Galactic emission, we selected $\beta = 2$ to reproduce the widths of the formal probability DPDF.  To verify the validity of this choice, we computed the GRS distance matching success rate (see \S\ref{res:grs_comp}) as a function of $\beta$, and found no dependence on the width of DPDF$_\mathrm{emaf}$.

The resulting DPDFs for object G035.524-00.274 (BGPS \#5647) are shown in Figure~\ref{fig:morph}{\it d}.  The prior DPDF$_{\mathrm{H}_2}$ (blue dot-dashed) favors the near kinematic distance, since this line of sight looks out the bottom of the molecular layer.  The gray dotted line shows the \ffore(\dsun) from the numerical model.  The morphological matching process, represented by DPDF$_\mathrm{emaf}$ (black solid), could not make the synthetic image dark enough to match the smoothed GLIMPSE image, forcing the prior to peak at \dsun~=~0~kpc.  The posterior DPDF (red) clearly reflects the distance selection, with the near peak containing $> 95\%$ of the integrated probability, although there remains some probability contained in the far kinematic distance peak at \dsun~$\approx 11$~kpc.


\section{Results}\label{sec:results}

\subsection{EMAF-Selected Molecular Cloud Clumps}\label{res:sources}

\subsubsection{Spatial and Kinematic Selection Criteria}

\begin{figure*}[t]
        \centering
        \includegraphics[width=6.5in]{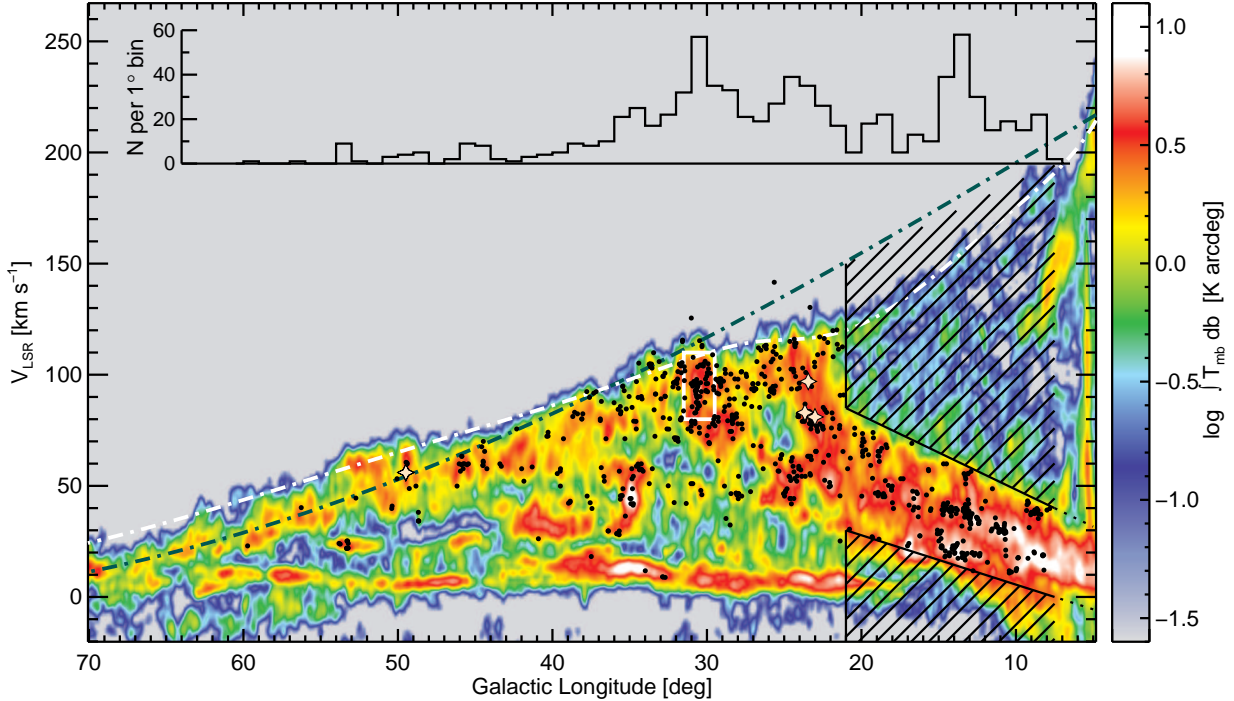}
        \caption{Longitude-velocity diagram of the northern Galactic plane.  The background image is the latitude-integrated ($|b| \leq 2\degr$) \twco(1-0) intensity from \citet[][]{Dame:2001}.  Excluded regions, where the long Galactic bar causes significant deviations from circular motion, are hashed out.  Black circles mark the locations of BGPS sources used in this study; the histogram at the top summarizes their Galactic longitude distribution.  Stars show the locations of masers used for parallax distance comparison (see Table~\ref{table:train}).  The white rectangle at $(\ell,$\vlsr$)\approx(30\degr,100$~\kms) roughly marks the W43 star-formation region (see \S \ref{disc:discrim}).  Colored dot-dashed lines represent the \citet[][white]{Clemens:1985} and \citet[][dark green, includes counter-rotation of HMSFRs]{Reid:2009} rotation curve tangent velocities as a function of Galactic longitude.  (A color version of this figure is available in the online journal.)}
        \label{fig:lv_co}
\end{figure*}

We derived posterior DPDFs for the subset of BGPS sources that have a measured \vlsr\ from molecular spectroscopy, and are selected by the presence of an EMAF.  Spatially, this set is defined by the GLIMPSE-BGPS overlap, limiting the upper end of the Galactic plane at $\ell = 65\fdg25$ and a latitude spread of $|b| \leq 1\fdg0$.  Kinematic considerations restrict the regions of the $\ell-v$ diagram (Fig.~\ref{fig:lv_co}) that may be considered at lower Galactic longitude down to the $\ell=7\fdg5$ limit of the spectroscopic surveys.  

The colored image in Figure~\ref{fig:lv_co} is the latitude-integrated \twco(1-0) intensity from \citet[][]{Dame:2001}, and is shown as an indicator of molecular gas location and kinematic conditions.  The presence of a long Galactic bar \citep[$R_\mathrm{bar}\sim4$~kpc;][]{Benjamin:2005} implies significant non-circular motion at $\ell \lesssim 30\degr$.  Regions at these longitudes in the $\ell-v$ diagram associated with the Molecular Ring feature \citep[\cf][]{Dame:2001,RodriguezFernandez:2008}, however, are likely at $R_\mathrm{gal} \gtrsim 4$~kpc.  To include the Ring but exclude bar-related gas, we disallowed the two hashed regions in Figure~5.  The upper region is bounded by \vlsr~= (3.33~\kms)$\times\ell(\degr) + 15$~\kms, and includes the higher-velocity gas inside the Ring.  The lower region excludes the 3-kpc expanding arm, and is bounded by \vlsr~= (2.22~\kms)$\times\ell(\degr) - 16.7$~\kms.  Both regions are defined only for $\ell \leq 21\degr$.  The upper hashed region does not extend past this point because the Molecular Ring feature extends to the tangent velocity at larger longitudes; the lower region is limited because the 3-kpc arm has its tangency here \citep{Dame:2008}.  There is likely overlap between Ring objects with nearly circular motions and objects in bar-related streaming orbits at $21\degr\leq\ell\leq30\degr$, so kinematic distance estimates in this range, including those derived here, should be used with caution.

Black circles mark the locations of the BGPS molecular cloud clumps for which a DPDF$_\mathrm{emaf}$ was computed, and the histogram summarizes their longitude distribution.  Stars mark the masers used for distance comparison (see \S\ref{res:maser_comp}).

\subsubsection{Mid-Infrared Selection Criteria}

Automated classification of dust-continuum-identified molecular cloud clumps as EMAFs was achieved using the mid-infrared contrast computed from the smoothed GLIMPSE images at the location of the BGPS source.  The peak contrast was defined as
\beqn\label{eqn:contrast}
C = 1 - \frac{I_\mathrm{min}}{\langle I_{_\mathrm{MIR}} \rangle}\ ,
\eeqn
where the intensity values were measured from the processed postage-stamp images described in \S\ref{sec:ubc_proc} for each Bolocat object.  Due to the varied sizes and shapes of EMAFs, standardized intensities were measured in a 40\arcsec\ aperture around the location of peak BGPS flux density (pink circles in Fig.~\ref{fig:glimpse_ubc}).  The value of $I_\mathrm{min}$ is the minimum intensity within the aperture measured from the smoothed star-subtracted GLIMPSE image (Fig.~\ref{fig:glimpse_ubc}{\it c}), and $\langle I_{_\mathrm{MIR}} \rangle$ is the mean of the $I_{_\mathrm{MIR}}$ postage-stamp image within 2\arcmin\ of the peak of millimeter flux density (cyan circle in Fig.~\ref{fig:glimpse_ubc}{\it d}).  A preliminary contrast threshold of $C \geq 0.01$ was implemented in the automated source selection to minimize the number of spurious matches caused by unrelated variation in the GLIMPSE 8-\micron\ mosaics.  This threshold also rejects BGPS sources that are mid-infrared bright, as those objects have negative contrast.


\begin{deluxetable*}{cccccccccccc}
  \tablecolumns{12}
  \tablewidth{0pc}
  \tabletypesize{\scriptsize}
  \tablecaption{Observed \& Derived Properties of EMAF-selected BGPS Molecular Cloud Clumps\label{table:sources}}
  \tablehead{
    \multicolumn{4}{c}{BGPS V1.0 Catalog Properties\tablenotemark{a}} & 
    \colhead{} & \multicolumn{2}{c}{Velocity Data} & \colhead{} & \colhead{} \\
    \cline{1-4} \cline{6-7}
    \colhead{Catalog} & \colhead{$\ell$} & \colhead{$b$} &
    \colhead{$S_{40}$\tablenotemark{b}} & \colhead{} &
    \colhead{\vlsr} & \colhead{Ref.} &
    \colhead{Mid-Infrared} &
    \colhead{KDA\tablenotemark{c}} & \colhead{\pchoose\tablenotemark{d}} &
    \colhead{\dml\tablenotemark{e}} &\colhead{\dbar~\tablenotemark{f}} \\
    \colhead{Number} & \colhead{(\degr)} & \colhead{(\degr)} &
    \colhead{(Jy)} & \colhead{} & \colhead{(\kms)} & \colhead{} &
    \colhead{Contrast} &
    \colhead{Resol.} & \colhead{} & \colhead{(kpc)} & \colhead{(kpc)}
  }
  \startdata
  4638 & 30.990 & 0.329 & 0.186(0.048) & & 79.1 & 2 & 0.21(0.03) & N & 0.88 & $~4.60^{+0.44}_{-0.40}$ & \nodata \\
  4639 & 30.990 & 0.385 & 0.108(0.054) & & 78.7 & 2 & 0.08(0.03) & F & 0.91 & $~9.88^{+0.38}_{-0.42}$ & \nodata \\
  4650 & 31.016 & -0.001 & 0.280(0.057) & & 74.5 & 1 & 0.25(0.04) & N & 0.88 & $~4.46^{+0.42}_{-0.42}$ & \nodata \\
  4653 & 31.026 & -0.113 & 0.289(0.067) & & 76.8 & 1 & 0.40(0.03) & N & 0.83 & $~4.50^{+0.52}_{-0.48}$ & \nodata \\
  4655 & 31.032 & 0.783 & 0.267(0.110) & & 51.0 & 1 & 0.38(0.05) & N & 0.99 & $~3.24^{+0.36}_{-0.36}$ & \nodata \\
  4715 & 31.226 & 0.023 & 0.381(0.077) & & 74.5 & 1 & 0.40(0.02) & N & 0.86 & $~4.40^{+0.48}_{-0.48}$ & \nodata \\
  4749 & 31.342 & -0.149 & 0.111(0.048) & & 42.0 & 1,3 & 0.15(0.03) & N & 0.91 & $~2.84^{+0.44}_{-0.44}$ & \nodata \\
  4769 & 31.432 & 0.167 & 0.117(0.039) & & 101.7 & 3 & 0.06(0.05) & U & 0.51 & \nodata & \nodata \\
  4770 & 31.436 & -0.103 & 0.122(0.039) & & 89.4 & 1,3 & 0.18(0.02) & U & 0.70 & \nodata & \nodata \\
  4780 & 31.462 & 0.351 & 0.090(0.046) & & 97.3 & 3 & 0.09(0.01) & N & 0.82 & $~5.56^{+0.72}_{-0.56}$ & \nodata \\
  4781 & 31.466 & 0.185 & 0.255(0.054) & & 103.6 & 3 & 0.19(0.08) & U & 0.70 & \nodata & \nodata \\
  4794 & 31.516 & 0.449 & 0.184(0.048) & & 83.7 & 3 & 0.11(0.04) & N & 0.91 & $~4.94^{+0.42}_{-0.40}$ & \nodata \\
  4811 & 31.580 & 0.227 & 0.186(0.048) & & 115.7 & 1,3 & 0.14(0.06) & T & 0.57 & $~7.00^{+0.78}_{-0.62}$ & $7.13(0.66)$ \\
  4814 & 31.584 & 0.205 & 0.218(0.052) & & 114.9 & 3 & 0.15(0.04) & T & 0.61 & $~6.88^{+0.82}_{-0.56}$ & $7.07(0.66)$ \\
  4826 & 31.608 & 0.171 & 0.120(0.042) & & 105.4 & 3 & 0.10(0.03) & U & 0.60 & \nodata & \nodata
  \enddata
  \tablenotetext{a}{\citet{Rosolowsky:2010}}
  \tablenotetext{b}{Flux density and uncertainty within a 40\arcsec\ aperture, corrected by the factor of $1.5\pm0.15$ from \citet{Aguirre:2011}}
  \tablenotetext{c}{N = near; F = far; T = tangent point; U = unconstrained distance}
  \tablenotetext{d}{Integrated posterior DPDF on the side of \dtan\ containing \dml.  Larger values indicate higher certainty in the KDA resolution.}
   \tablenotetext{e}{Maximum-likelihood distance; the distance where the posterior DPDF is largest.  Not listed for unconstrained sources.}
   \tablenotetext{f}{Weighted-average distance; the first moment of the posterior DPDF.  Only listed for sources at the tangent point.}
  \tablerefs{1: \hcop \citep{Shirley:2013}; 2: CS (Y. Shirley 2012, private communication); 3: \nhhh\ \citep{Dunham:2011c}}
  \tablecomments{Errors are given in parentheses.}
  \tablecomments{Table \ref{table:sources} is published in its entirety in a machine-readable format in the online journal.  A portion is shown here for guidance regarding its form and content.}
\end{deluxetable*}

All molecular cloud clumps meeting the above selection criteria were examined by eye to ensure their suitability for deriving a DPDF$_\mathrm{emaf}$.  Bolocat objects were not assigned a DPDF$_\mathrm{emaf}$ for the following types of deficiencies: (1)~there was evidence of poor star subtraction contaminating $I_\mathrm{min}$; (2)~there was very bright mid-infrared emission ($I_{8\mu \mathrm{m}}$~$\gtrsim 200$~MJy~sr$^{-1}$) within 2\arcmin\ of the location of peak BGPS flux density that could bleed into the 40\arcsec\ aperture or significantly affect the IRAC scattering correction; (3)~the postage-stamp estimate of $I_{_\mathrm{MIR}}$ was contaminated by excessive bright emission or dark extinction; or (4)~the morphology of dark regions in the GLIMPSE image clearly did not correspond to that of the millimeter emission.  By-eye exclusion removed approximately 40\% of sources meeting the initial automated selection criteria.  

Properties of rejected sources were analyzed to reveal that nearly all very-low contrast sources were spurious matches (deficiency type 4, see above).  Additionally, BGPS objects located in fields of locally very bright mid-infrared emission were almost all excluded from the final source list (types 2 and 3).  As a result, the contrast cutoff was increased to $C \geq 0.05$, and two additional automated selection criteria were introduced.  First, the restriction $\langle I_{_\mathrm{MIR}} \rangle \leq 100$~MJy~sr$^{-1}$ was placed to remove sources whose background estimate indicates significant disagreement with the 8-\micron-emission model, as large discrepancies may lead to improper distance estimates (type 3).  Second, to automatically reject sources near very bright emission, the re-pixelated unsmoothed postage-stamp images were checked for pixels with $I_{8\mu \mathrm{m}} \geq 200$~MJy~sr$^{-1}$ within 2\arcmin\ of the image center; sources with more than 10 such (7\farcs2) pixels were removed from consideration (type 2).  These additions to the automated selection criteria led to fewer sources (only 28\%) requiring by-eye removal, primarily due to poor star-subtraction (type 1) or complex emission structures that caused morphological mismatch (type 4).

\begin{figure*}[t]   
        \centering
        \includegraphics[width=2.1in]{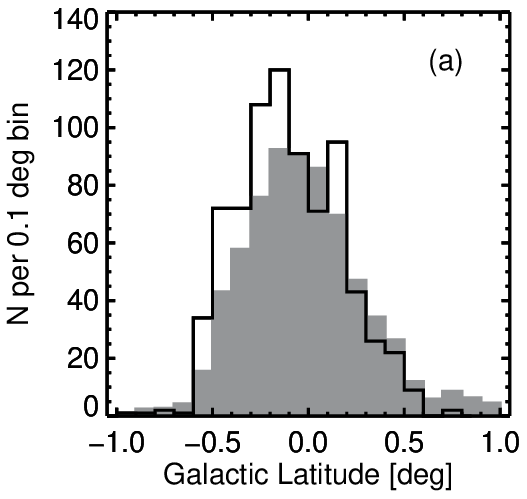}
        \includegraphics[width=2.1in]{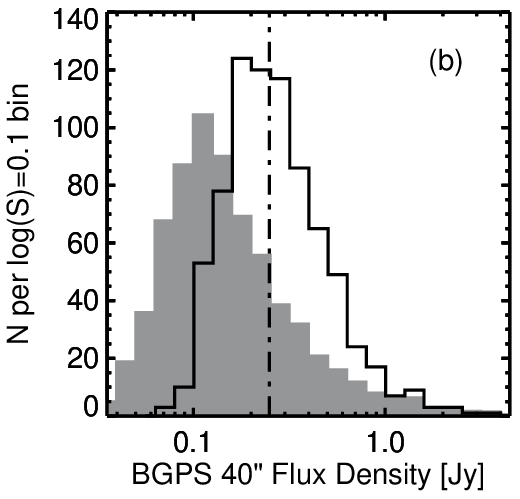}
        \includegraphics[width=2.1in]{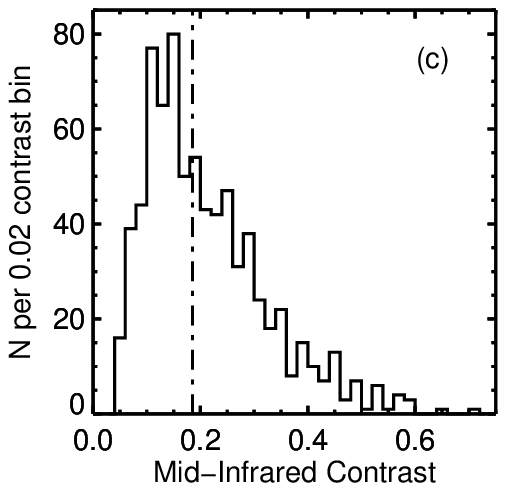}
        \caption{({\it a}) Distribution of Galactic latitude of the sources in this study (black outline) and the entire BGPS catalog in the longitude range $7\fdg5\leq\ell\leq65\degr$, divided by 10 (filled gray).  The BGPS is nominally limited to $|b| \leq 0.5$.  ({\it b}) Distribution of BGPS 40\arcsec\ flux density for the sources in this study (black outline) and the full, longitude-limited BGPS catalog, divided by 10 (filled gray).  Vertical dot-dashed line marks the median for this sample.  ({\it c}) Distribution of measured mid-infrared contrast for the sources in Table~\ref{table:sources}.  Vertical dot-dashed line marks the median.}
        \label{fig:source_hist1}
\end{figure*}

\subsubsection{Source Properties}

The final source list contains 770 BGPS objects, and is presented in Table~\ref{table:sources}.  EMAF-selected BGPS molecular cloud clumps are not drawn uniformly from the BGPS catalog.  Comparisons of Galactic latitude and 40\arcsec\ flux density distributions between this sample and the full Bolocat (within the spatial limits defined above) are shown in Figure~\ref{fig:source_hist1}.  The latitude distribution of this sample follows that of the BGPS as a whole, including peaking below $b=0\degr$.  The offset is related to the Sun's vertical displacement above the Galactic midplane \citep{Schuller:2009,Rosolowsky:2010}.  The only significant deviation is near $b = 0\degr$, where locally bright 8-\micron\ emission along the midplane, excited by \HII\ regions and OB stars, obscures more distant molecular cloud clumps.  The BGPS 40\arcsec\ flux density histograms (Fig.~\ref{fig:source_hist1}{\it b}) show that this sample contains, on average, brighter sources (median = 0.252~Jy) than the full Bolocat (median = 0.135~Jy).  There are two likely origins of this bias.  First, sources must have a \vlsr\ measurement from a dense-gas tracer; the \hcop\ detection fraction, in particular, is a strong function of BGPS flux density \citep[$<20\%$ for $S_{_{1.1}} < 0.1$~Jy;][]{Shirley:2013}.  Second, the selection criteria excluded sources with very low contrast or whose morphology does not correspond to dark regions in the GLIMPSE maps.  Faint BGPS sources have low optical depth ($S_{_{1.1}} = 0.1$~Jy corresponds to $\tau_{_8} \approx 0.07$), and would be difficult to distinguish against the variable Galactic 8-\micron\ background.

The distribution of measured mid-infrared source contrast is shown in Figure~\ref{fig:source_hist1}{\it c}, and has a median of 0.19.  For images without the IRAC scattering correction, this value corresponds to an uncorrected median contrast of 0.15 (see Appendix \ref{app:scatter}), considerably lower than the minimum contrast ($C \approx 0.20$) used by \citet{Peretto:2009} in their catalog of \spitzer\ IRDCs (which did not correct for IRAC scattering in the same manner).  The majority of our sample consists of ``low-contrast'' sources that are missing from published catalogs of Galactic IRDCs.

\subsection{Source KDA Resolutions}\label{res:dist}

\subsubsection{Distance Estimates and Constraints}\label{res:constr}

\begin{figure}[t]   
        \centering   
        \includegraphics[width=3.1in]{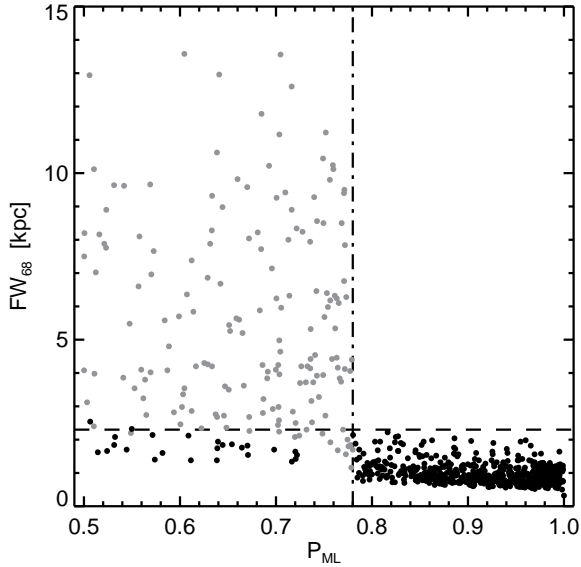}
        \caption{Comparison of the full-width of the 68.3\% error bar (FW$_{68}$) against the integrated DPDF on the \dml\ side of \dtan\ (\pchoose).  The vertical dot-dashed line represents the empirical cutoff at \pchoose~=~0.78, and the horizontal dashed line marks FW$_{68}$~=~2.3~kpc.  Objects shown in gray are below the \pchoose\ cutoff and are more than 1~kpc from the tangent point.  We defined FW$_{68} \leq 2.3$~kpc as the criterion for a ``well-constrained'' distance estimate, which encompasses the objects in the bottom left corner, as well.}
        \label{fig:constrain}
\end{figure}

We derived posterior DPDFs for the EMAF-selected BGPS sources by multiplying the kinematic distance DPDF by the two priors, and normalizing to unit total probability.  By design, the prior DPDFs are broader than the peaks in DPDF$_\mathrm{kin}$, so the resulting maximum-likelihood distances generally do not differ from the simple kinematic distances by more than $\sim 0.1$~kpc.  To gauge the strength of the KDA resolution, two statistics were defined: the maximum-likelihood probability (\pchoose) as the integrated posterior DPDF on the \dml\ side of the tangent point, and the full width of the 68.3\% maximum-likelihood error bar (FW$_{68}$).  The ranges of these statistics are 0.5~$\leq$ \pchoose~$\leq 1.0$ and 0.2~kpc~$\lesssim$ FW$_{68}$~$\lesssim$ 15~kpc, and a comparison between them for each object is shown in Figure~\ref{fig:constrain}.  The nature of a double-peaked DPDF$_\mathrm{kin}$ leads to the sharp change in the distribution of FW$_{68}$ near \pchoose~=~0.78 (vertical dashed line).  When the ratio of the peak probabilities of the kinematic distance peaks in the posterior DPDF becomes $\lesssim 3$, the error bars must include both to enclose sufficient probability.  For objects with \pchoose~$\geq 0.78$, the maximum-likelihood error bars enclose a single kinematic distance peak.  We consider this set to have well-constrained distance estimates, and note that  FW$_{68} \leq 2.3$~kpc (horizontal dashed line).  Objects with full-width error bars less than this value and are below the \pchoose\ cutoff (lower-left corner of Fig.~\ref{fig:constrain}) are generally within $\sim 1$~kpc of the tangent distance.  Because of the limited distance range available to these sources, their distance estimates should also be considered well-constrained.  Combining these sets, we adopted FW$_{68} \leq 2.3$~kpc as the criteria for well-constrained distance estimates.

Objects within a kiloparsec of \dtan\ have posterior DPDFs that are oftentimes asymmetric, and \dml\ is not the best single-value representation of the distance.  For these objects, we assigned them to the tangent distance group, and used \dbar\ (and associated uncertainty) in the analysis that follows.  A total of 618 sources in this sample have well-constrained distance estimates (80\%).  The KDA resolutions for these objects are recorded as ``N'' (near), ``F'' (far), or ``T'' (tangent distance) in column~8 of Table~\ref{table:sources}; \dml\ is listed in column~10.  The remaining objects are recorded as ``U'' (unconstrained) and have no distance estimate listed.  For the tangent group, the weighted-average distance (\dbar) is listed in column~11, and is the preferred distance representation for these objects (\dml\ is shown for comparison only).  DPDFs for all sources are available in the BGPS archive.

\subsubsection{Heliocentric Distances}\label{res:dist_helio}

\begin{figure*}[t]
        \centering
        \includegraphics[width=2.1in]{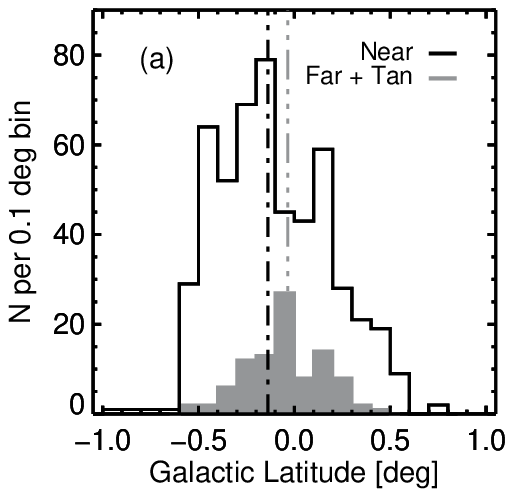}
        \includegraphics[width=2.1in]{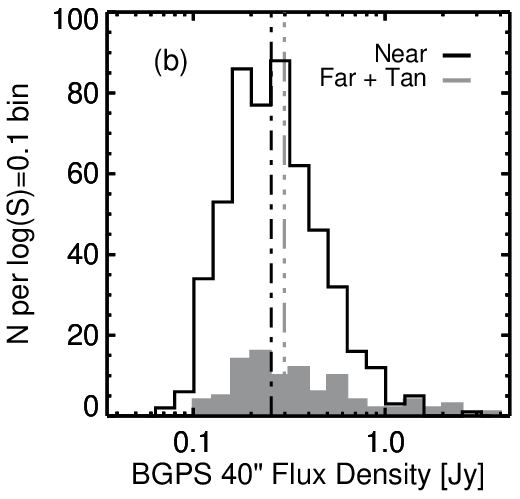}
        \includegraphics[width=2.1in]{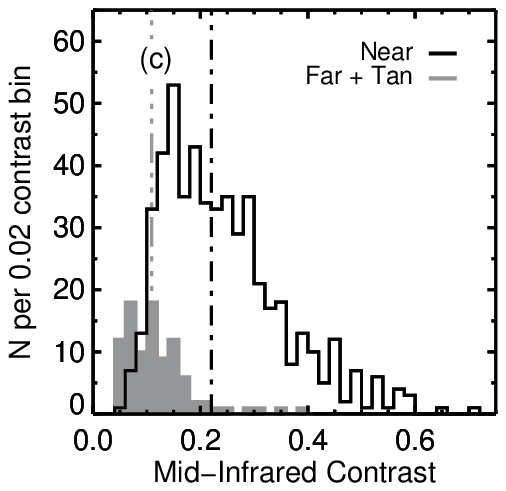}
        \caption{Comparison of source properties for objects with ``near'' vs. ``far'' KDA resolutions.  Sources placed at the near distance are represented by open black histograms; tangent-point and far-distance sources are shown with filled gray histograms.  Panels are as in Fig.~\ref{fig:source_hist1}.  Vertical dot-dashed lines represent the median of each distribution.}
        \label{fig:nf_hists}
\end{figure*}

For the remaining discussion, we consider only the 618 mid-infrared-dark BGPS sources whose distances are well-constrained (as defined above).  Of this set, 70 were placed beyond the tangent point, with another 25 near \dtan, indicating the significant possibility of detecting molecular cloud clumps at the far kinematic distance using mid-infrared absorption.  The comparisons of latitude distribution, BGPS 40\arcsec\ flux densities, and mid-infrared contrast between the near and far subsets are shown in Figure~\ref{fig:nf_hists}.  For the purposes of this discussion, objects at the tangent distance are grouped with those at the far kinematic distance.  The latitude distributions (panel~{\it a}) are very similar, with the near group being slightly wider, owing to the latitude-limiting effect of DPDF$_{\mathrm{H}_2}$.  

The histograms of BGPS 40\arcsec\ flux densities (Fig.~\ref{fig:nf_hists}{\it b}) show that the far subset has a flatter distribution with a higher median than the near set.  Since the source list for this study is mid-infrared-contrast limited, we do not expect to see low flux-density BGPS sources at the far distance; the low column density would not produce enough attenuation to be seen behind the significant foreground emission.  The expected contrast as a function of heliocentric distance (represented by \ffore) may be computed by combining Equations (\ref{eqn:imin_model}), (\ref{eqn:tau_8}), and (\ref{eqn:contrast}) into
\beqn\label{eqn:cfs}
C = (1 - f_\mathrm{fore})(1 - e^{-\Upsilon S_{_{1.1}}})\ ,
\eeqn
where $I_\mathrm{emaf}$ and $I_{_\mathrm{MIR}}$ from Equation (\ref{eqn:imin_model}) are equivalent to $I_\mathrm{min}$ and $\langle I_{_\mathrm{MIR}} \rangle$ from Equation (\ref{eqn:contrast}), respectively.  A source with larger flux density may be at a farther \dsun\ and still meet the contrast selection criterion.

Indeed, sources at the far kinematic distance have a lower median mid-infrared contrast ($C = 0.11$) than those at the near kinematic distance ($C = 0.22$), as shown in Figure~\ref{fig:nf_hists}{\it c}.  Of the 313 objects with $C \geq 0.2$, only 12 (4\%) are placed at the far kinematic distance, reinforcing the notion that dark mid-infrared absorption features must lie relatively nearby.  For comparison, of the 63 sources with $C < 0.1$, 40 (63\%) were placed at the far kinematic distance, indicating that the majority of EMAFs with very low contrast are at or beyond the tangent point.

\begin{figure}[t]
        \centering
        \includegraphics[width=3.1in]{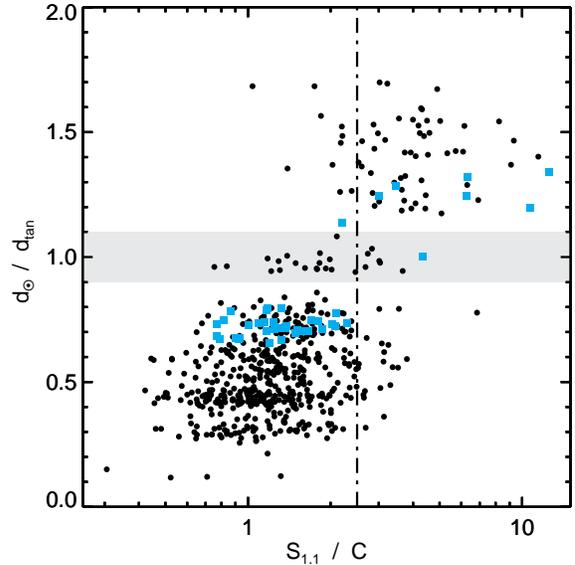}
        \caption{KDA resolution versus the ratio of BGPS 40\arcsec\ flux density to mid-infrared contrast.  The upper region represents the far kinematic distance, and the lower the near; the gray shaded region illustrates the band around \dtan.  Black dots mark the sources with well-constrained distance estimates; the subset of cyan squares are sources within W43 (see \S\ref{disc:discrim}).  The vertical dot-dashed line is drawn at $S_{_{1.1}} / C = 2.5$.  (A color version of this figure is available in the online journal.)}
        \label{fig:sc_ratio}
\end{figure}

An interesting empirical predictor of KDA resolution is shown in Figure~\ref{fig:sc_ratio}.  The ratio of resolved heliocentric distance over \dtan\ is plotted against the ratio of BGPS 40\arcsec\ flux density over the mid-infrared contrast. For BGPS data, there appears to be a boundary at $S_{_{1.1}} / C \approx 2.5$ that divides KDA resolutions.  The exact value of this cutoff is dependent upon the (sub-)millimeter survey used, and there exists scatter across the boundary.  It nevertheless suggests an additional means for KDA resolution when DPDF$_\mathrm{emaf}$ fails to return a well-constrained estimate.

\subsubsection{Galactocentric Positions}\label{res:dist_gal}

\begin{figure}[t]
        \centering
        \includegraphics[width=3.1in]{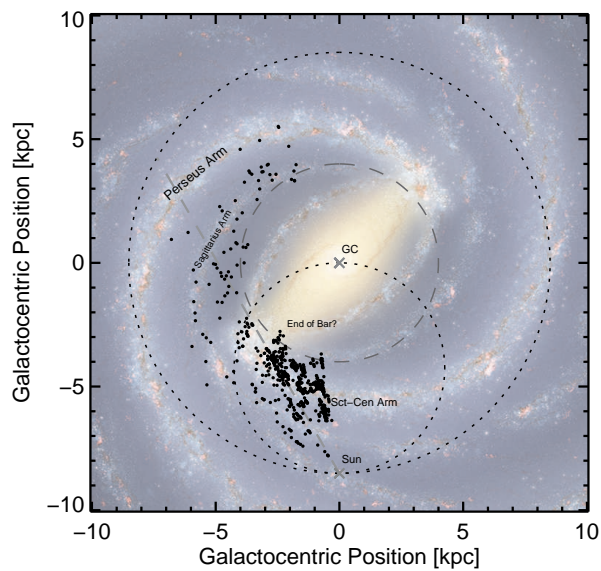}
        \caption{Face-on view of the Milky Way for sources with well-constrained KDA resolutions, plotted atop an artist's rendering of the Milky Way (R. Hurt: NASA/JPL-Caltech/SSC) viewed from the north Galactic pole.  The image has been scaled to match the $R_0$ used for calculating kinematic distances.  The outer dotted circle marks the Solar circle, and the inner dotted circle the tangent point as a function of longitude.  The dashed circle at $R_\mathrm{gal} = 4$~kpc outlines the region influenced by the long Galactic bar \citep{Benjamin:2005}, corresponding to the hashed regions in Fig.~\ref{fig:lv_co}.  The straight dashed gray line marks $\ell=30\degr$ as a guide.  Various suggested Galactic features are labeled.  For clarity, distance error bars are not shown.  (A color version of this figure is available in the online journal.)}
        \label{fig:galpos2}
\end{figure}

With well-constrained distance estimates, it is possible to construct a face-on view of the Milky Way.  Sources with well-constrained KDA resolutions are plotted atop a reconstruction of the Milky Way from \spitzer\ data in Figure~\ref{fig:galpos2} (R. Hurt: NASA/JPL-Caltech/SSC) using either \dml\ or \dbar\ as described in \S\ref{res:constr}.  For clarity, the error bars, which account for small deviations from circular motion, are not shown.  Some spiral structure is evident in the map of BGPS sources, notably portions of the Sagittarius Arm at $\ell \gtrsim 35\degr$, the Scutum-Centarus Arm / Molecular Ring at $\ell \lesssim 30\degr$, and the local arm / Orion spur within about a kiloparsec of the Sun \citep{Churchwell:2009}.  The kinematic restrictions on our sample led to the absence of objects within a $\approx 3.5$~kpc radius of the Galactic center (dashed circle in the figure).  Face-on views of the Galaxy derived from kinematic distances will not show narrow spiral features (like those in the background image) because of the local virial motions of individual molecular cloud clumps within larger complexes.  Galactocentric positions are therefore ``smeared-out'' by approximately $\pm 0.4$~kpc about the true kinematic distance for the complex as a whole.  Each dot in the figure, however, should be thought of in terms of its DPDF, where the kinematic distance peaks have a FWHM of 1-2~kpc.

\begin{figure}[t]
        \centering
        \includegraphics[width=3.1in]{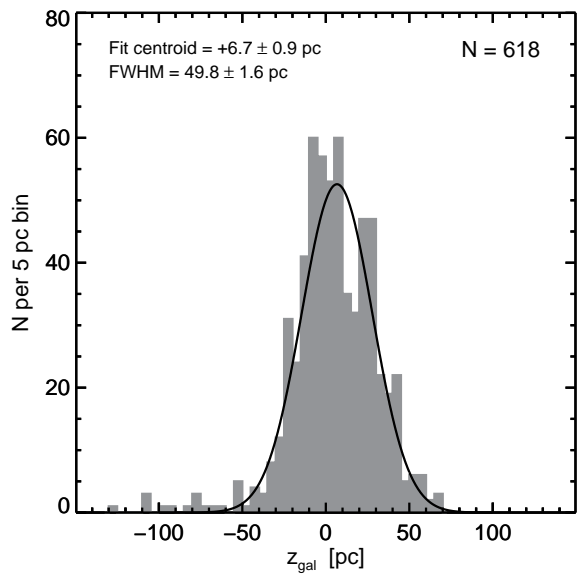}
        \caption{Vertical distribution of sources about the Galactic midplane.  The filled gray histogram shows the distribution, while the black line represents a Gaussian fit to the histogram.}
        \label{fig:z_pos}
\end{figure}

\begin{figure}[t]
        \centering
        \includegraphics[width=3.1in]{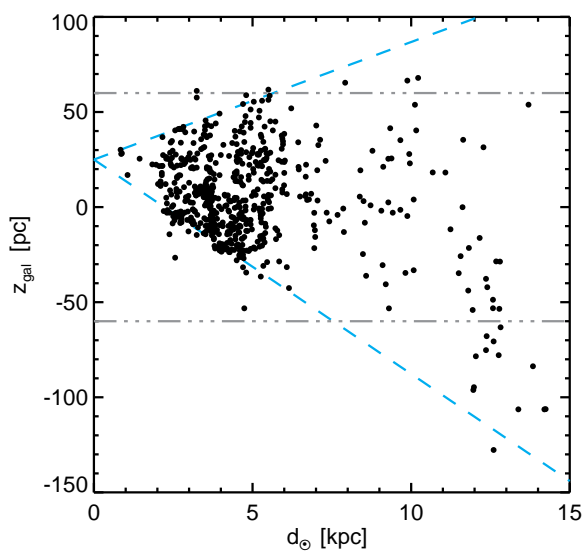}
        \caption{Derived vertical position of sources versus heliocentric distance.  Diagonal cyan dashed lines represent the nominal $|b| = 0\fdg5$ limit of the BGPS at $\ell = 30\degr$ for a vertical Solar offset above the Galactic midplane of 25~pc.  Horizontal dot-dashed lines mark the FWHM of the \twco\ layer \citep{Bronfman:1988}.  (A color version of this figure is available in the online journal.)}
        \label{fig:z_dsun}
\end{figure}

\begin{deluxetable*}{lcccccc}
  \tablecolumns{7}
  \tablewidth{0pc}
  \tabletypesize{\small}
  \tablecaption{Maser Sources for Distance Comparison\label{table:train}}
  \tablehead{
    \colhead{Source} & \colhead{$\ell$} & 
    \colhead{$b$} & \colhead{\vlsr} & \colhead{Distance} & 
    \colhead{$N_{_\mathrm{BGPS}}$\tablenotemark{a}} & \colhead{Ref.} \\
    \colhead{Name} & \colhead{(\degr)} & 
    \colhead{(\degr)} & \colhead{(km s$^{-1}$)} & \colhead{(kpc)} & 
    \colhead{} & \colhead{}
  }
  \startdata
  G23.0-0.4 & 23.01 & -0.41 & 81.5 & $ 4.6^{+0.4}_{-0.3}  $ & 6 & 1 \\
  G23.4-0.2 & 23.44 & -0.18 & 97.6 & $ 5.9^{+1.4}_{-0.9 } $ & 2 & 1 \\
  G23.6-0.1 & 23.66 & -0.13 & 82.6 & $ 3.2^{+0.5}_{-0.4}  $ & 2 & 2 \\
  W51 IRS2  & 49.49 & -0.37 & 56.4 & $ 5.1^{+2.9}_{-1.4}  $ & 1 & 3 \\
  W51 Main\tablenotemark{b}  & 49.49 & -0.39 & 58.0 & $ 5.4^{+0.3}_{-0.3}  $ & 1 & 4
  \enddata
  \tablenotetext{a}{Number of EMAF-selected BGPS sources within 15\arcmin\ and 10~\kms\ of the maser location.  See Fig. \ref{fig:dist_comp} for the comparison.}
  \tablenotetext{b}{H$_2$O maser; all others are CH$_3$OH masers}
  \tablerefs{(1) \citet{Brunthaler:2009}; (2) \citet{Bartkiewicz:2008}; (3) \citet{Xu:2009}; (4) \citet{Sato:2010}}
\end{deluxetable*}

KDA resolutions also allow the derivation of the vertical distribution of sources about the Galactic midplane.  Vertical position is particularly affected by the KDA for higher-latitude sources ($|b| \gtrsim 0\fdg4$).  Calculation of vertical height ($z$) requires a proper accounting of the Sun's $\approx 25$~pc vertical offset above the Galactic plane \citep{Humphreys:1995,Juric:2008}.  The small scale height of Galactic molecular gas can lead to incorrect inferences about the vertical distribution of dense gas in the disk if  $z$ positions are calculated directly from Galactic coordinates without a correction for the Solar offset.  The matrix needed to transform ($\ell,b,d$) into ($R_\mathrm{gal}, \phi, z$) is derived in Appendix~\ref{app:coord_conv}.

The vertical distribution for the set of well-constrained BGPS sources is shown in Figure~\ref{fig:z_pos}.  A Gaussian fit to the distribution yields a half-width at half maximum of 25~pc, and a positive centroid offset of 7~pc.  This scale height is approximately half that found by \citet{Bronfman:1988} for \twco.  This narrow result may, however, be a result of the limited Galactic latitude coverage of the BGPS.  Analysis of the recent compact source catalog from the $\lambda = 870$~\micron\ ATLASGAL survey \citep{Contreras:2013}, which extends to $|b| = 1\degr$, shows that $\approx 20\%$ of their objects lie outside the BGPS latitude limits.  To gauge the effect of limited latitude coverage, the derived $z$ are plotted against heliocentric distance in Figure~\ref{fig:z_dsun}, with $|b| = 0\fdg5$ shown for $\ell = 30\degr$ (the limits rotate to slightly more positive $z$ for larger $\ell$).  For the region \dsun~$\leq 6$~kpc (which contains more than 80\% of this sample), the BGPS does not fully probe the FWHM of the \twco\ distribution (dot-dashed lines).  Other indicators of a larger scale height for star-formation regions include \thco\ clouds from the GRS \citep{RomanDuval:2009}, which have a FWHM~$\approx 80$~pc, and Galactic \HII\ regions \citep[FWHM~$\approx 100$~pc;][]{Anderson:2012}.

\subsection{Distance Comparisons with Other Studies}\label{res:comps}

The quality of KDA resolutions for EMAF-selected BGPS sources was characterized through a comparison of distance estimates with values from the literature.  In particular, comparison sets were chosen that used mostly orthogonal methodologies so that distance comparisons are largely free of correlated effects.  The three sets described below are the use of maser parallax measurements, \HI\ absorption features associated with molecular clouds, and near-infrared extinction measurements.

\subsubsection{Maser Parallax Distances}\label{res:maser_comp}

\begin{figure*}[t]
        \centering
        \includegraphics[width=3.1in]{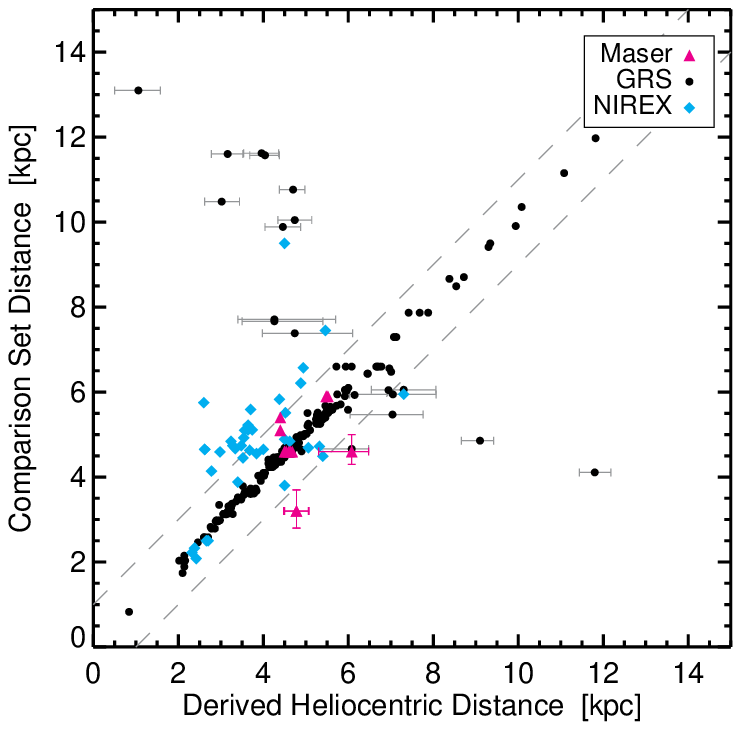}
        \includegraphics[width=3.1in]{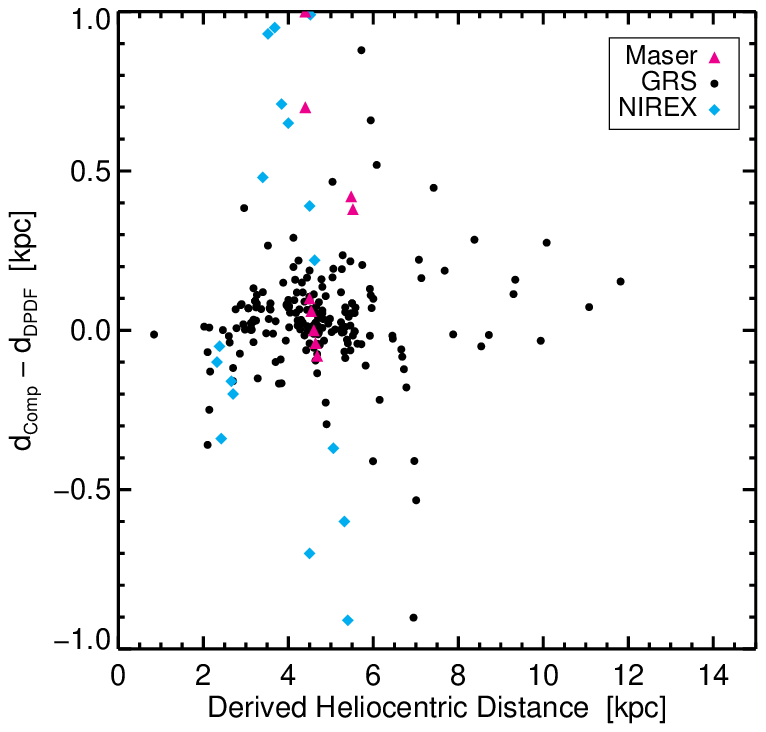}
        \caption{{\it Left:} Comparison of KDA resolutions derived from the DPDF with published distance estimates.  Gray dashed lines represent $\pm 1$~kpc away from equality.  Maximum-likelihood horizontal error bars are shown for GRS- and maser-associated sources lying outside this region.  Vertical error bars for maser-associated sources come from Table~\ref{table:train}.  See text for discussion of systematic offset for NIREX sources.  {\it Right:} A zoom-in on the region $\pm 1$~kpc from distance equality to better visualize the distance comparison for these objects.  The vertical axis represents the comparison set distance minus the heliocentric distance from the DPDF.  (A color version of this figure is available in the online journal.)}
        \label{fig:dist_comp}
\end{figure*}

Maser parallax measurements towards regions of high-mass star formation provide absolute distance validation comparisons.  The Bar and Spiral Structure Legacy Survey \citep[BeSSeL;][]{Brunthaler:2011} is conducting ongoing VLBI parallax measurements of CH$_3$OH and H$_2$O maser emission in star-forming regions across the Galactic plane.  Such measurements provide very accurate distances out to \dsun~$\sim 10$~kpc, but the present overlap between published results and the BGPS is small (see Table~\ref{table:train} for the comparison set of maser sources used).  

The comparison set was defined as objects from our sample whose angular separations and velocity differences were $\leq 15\arcmin$ and $\leq 10$~\kms, respectively, from those of a published maser.  These masers tend to be in regions of high-mass star formation, and such regions are on order $0\fdg25$ in size.  The velocity limits are related to the spread of virial velocities within such regions.  A collection of 12 BGPS objects were associated with one of five masers; the distribution is noted in Table~\ref{table:train}.  To visualize the distance comparison, distances from Table~\ref{table:sources} for each BGPS object are plotted against the measurements from the BeSSeL literature as magenta triangles in  Figure~\ref{fig:dist_comp}.  The gray dashed lines in the left panel represent $\pm 1$~kpc error margins, used for qualitative purposes.  An object falling outside this region is said to have a ``mismatching'' distance estimate.  For clarity in the figure, only mismatching objects have error bars shown; horizontal bars are from Table~\ref{table:sources}, and vertical bars are from the reference.  The right panel is a zoom-in on the $\Delta d = \pm 1$~kpc region of the left panel (\ie within the dashed lines).

For the maser comparison set, only three sources fall outside the $\pm 1$~kpc region.  Two BGPS objects (overlapping triangles in Fig.~\ref{fig:dist_comp}) are associated with the maser source G23.66-0.13, which has a parallax distance that disagrees with the derived (near) kinematic distance.  \citet{Bartkiewicz:2008} find that this object has a proper motion consistent with the parallax distance and the assumption of a flat rotation curve, but has a $\approx 35$~\kms\ peculiar motion toward the Galactic center.  This radial streaming motion makes its kinematic distance appear larger, and provides a cautionary example of the effects of non-circular motion on kinematic distance methods.  The other mismatching source has a DPDF distance estimate skewed away from the simple kinematic distance due to the sharply-peaked DPDF$_\mathrm{emaf}$ caused by its bright millimeter flux density ($S_{_{1.1}} = 3.9$~Jy).  The W51 region lies near the tangent point, so correct DPDF distance placement for these objects merely implies that the region's circular velocity is consistent with the rotation curve.  The remaining EMAF-selected BGPS objects in this set have KDA resolutions that agree with the trigonometric parallax distance.

\subsubsection{Galactic Ring Survey KDA Resolutions}\label{res:grs_comp}

For a larger distance comparison set, we used the KDA resolutions from the BU-FCRAO Galactic Ring Survey \citep[GRS;][]{Jackson:2006}.  By matching \thco(1-0) emission morphology and spectra with \HI\ absorption features, \citet{RomanDuval:2009} estimated the distances to some 750 molecular clouds in the inner Galaxy.  Those authors used a combination of \HI\ self-absorption (HISA)\footnote{Absorption features caused by cold neutral hydrogen within molecular clouds are also called ``narrow'' self-absorption (HINSA) to distinguish them from the broader self-absorption features of diffuse \HI\ clouds \citep[\cf][]{Li:2003}.} and 21-cm continuum absorption features to positively resolve the KDA.  These techniques exploit the spectroscopic dimension of \HI\ surveys, where cold atomic hydrogen within dense molecular gas absorbs line emission from warm gas at the same \vlsr\ on the far side of the Galaxy or continuum radiation from \HII\ regions.  Distance resolutions from this method are subject to uncertainties from non-circular and radial streaming motions, but are directly comparable with the KDA resolutions of the DPDF method.

EMAF-selected BGPS objects were associated with cataloged \thco\ clouds based on spatial and kinematic proximity.  The association volume was defined as a circle of radius 10\arcmin\ \citep[approximately the median size of a GRS cloud;][]{RomanDuval:2009}, and a velocity spread equal to the \thco\ velocity dispersion, centered on the ($\ell,b,$\vlsr) coordinates from the GRS catalog.  A total of 213 EMAFs lie within the association volume of one or more GRS clouds.  To ensure the accuracy of the associated GRS KDA resolution,  BGPS flux density maps were compared to both \thco\ intensity maps integrated over the velocity of the appropriate dense-gas tracer from Table~\ref{table:molec}, and \HI\ 21-cm ``on''--``off'' integrated intensity (HISA) maps.  For a handful of sources ($\sim7\%$), a strong HISA signature was present within the BGPS source contour even though the associated GRS cloud was placed at the far kinematic distance.  None of these objects was listed as having a 21-cm continuum source, so the absorption signature is the result of cold gas at the near distance.  These discrepant objects may be the result of line-of-sight confusion, a slight velocity offset between the parent cloud and the BGPS object, or incorrect association with a \thco\ cloud.  Whatever the cause, the KDA resolution for the associated GRS cloud was amended to ``near'' to reflect the HISA signature.

\citeauthor{RomanDuval:2009} used the \citet{Clemens:1985} curve to derive heliocentric distances, and differences in rotation curve definition can cause distance-comparison discrepancies unrelated to the KDA (see Fig.~\ref{fig:lv_co}).  To eliminate potential systematic effects, the KDA resolution and \vlsr\ of each associated GRS cloud were mapped to a new heliocentric distance using the \citet{Reid:2009} rotation curve.  In the comparison between the GRS-derived distance and those from the DPDFs (black dots in Fig.~\ref{fig:dist_comp}), nearly 92\% of our distance resolutions match those of the GRS.  This success rate is robust for the entire EMAF set, as enforcing a minimum mid-infrared contrast of $C \geq 0.15$ only increases the matching rate to 94\%.

The 17 BGPS objects with mismatching distance resolutions are shown with horizontal error bars from the DPDFs.  Those in the upper-left of Figure~\ref{fig:dist_comp} have a large apparent mid-infrared absorbing column, but \citet{RomanDuval:2009} did not find evidence of self-absorbing \HI.  Conversely, those in the bottom-right have HISA signatures but were placed beyond the tangent point by the posterior DPDF.  Examination by eye of this latter group showed that the two sources farthest from the one-to-one line have slight underestimates of $I_{_\mathrm{MIR}}$ around the EMAF; the values in the postage-stamp image reflect dimmer nearby regions.  The DPDF$_\mathrm{emaf}$ in these cases selects the far kinematic distance peak despite the presence of HISA for these objects.

There are four objects whose GRS distance estimate is 5.5~kpc~$\geq$ \dsun~$\geq 8$~kpc and disagree with the DPDF-derived distance.  These all lie within $\sim 1.5$~kpc of the tangent point.  Since the kinematic distance DPDFs do not have two fully distinct peaks in this region, the particular shape of DPDF$_\mathrm{emaf}$ can have a significant impact on the derived single-distance estimators.  The mismatches are due to the source being near \dtan, and not an incorrect KDA resolution.  The remaining eleven mismatching sources in the upper-left of Figure~\ref{fig:dist_comp} (GRS-far, DPDF$_\mathrm{emaf}$-near) are moderately dark EMAFs ($0.1 \leq C \leq 0.3$) that show no signs of HISA at the velocity of the molecular cloud clump.  About half of these lie at $|b| \geq 0\fdg4$, and may not have enough \HI\ backlighting at the far kinematic distance for a HISA signature to be visible; although \citet{Gibson:2005a} found self-absorption features out to more than $|b| =2\degr$ in the Canadian (\HI) Galactic Plane Survey.  For sources in this quadrant of the figure, it is unclear which kinematic distance is correct.  The future application of additional prior DPDFs may solve the small number of conflicting KDA resolutions, but the present method achieves very good correspondence with other distance estimates for molecular cloud clumps.

\subsubsection{Near-Infrared Extinction Distances}

Using a technique for measuring three-dimensional near-infrared Galactic extinction \citep[NIREX;][]{Marshall:2006}, \citet{Marshall:2009} estimated the distances to over 1200 IRDCs identified by MSX in the inner Galactic plane \citep[][hereafter S06]{Simon:2006}.  This approach compares the stellar colors of a section of sky with a Galactic stellar distribution model, and searches for sharp changes in color excess as a function of distance.  Extinction measurements, like maser parallaxes, offer a kinematic-independent means of distance determination.

MSX dark clouds have a typical size of about an arcminute, so BGPS sources lying within 60\arcsec\ of the centroid of a NIREX cloud were included in this comparison set.  While there are about 275 objects from \citet{Marshall:2009} within the spatial bounds of this study, only 38 EMAF-selected BGPS sources could be associated with a NIREX cloud.  \citet[][hereafter PF09]{Peretto:2009} noted that only a quarter of MSX IRDCs appear in their catalog of \spitzer\ dark clouds for a variety of reasons.  This selection effect, in combination with our requirement that an object be detected in one or more molecular line transitions, makes the number of matching clouds reasonable.  

Objects from the NIREX comparison set are shown as cyan diamonds in Figure~\ref{fig:dist_comp}.  Most of the points lie within 2~kpc of equality.  Only one object has a wildly divergent distance estimate, G31.026-0.113 (BGPS \#4653; $d_{_\mathrm{NIREX}} \approx 9.5$~kpc), which has $C = 0.4$ and should have been detected with a strong near-infrared absorption signature at the near kinematic distance of \dsun~$=4.5$~kpc.  The collection of cyan diamonds with a systemic positive offset of 1.5~kpc warrants attention.  There is a cluster of objects placed 2-4~kpc from the Sun.  Most of these are at $\ell \lesssim 15\degr$, and uncertainties in both the rotation curve and stellar model in that region may be contributing to the offset.  Mismatching NIREX distances beyond \dsun~= 4~kpc have divergent distance estimates of order the difference between the \citet{Clemens:1985} and \citet{Reid:2009} rotation curves for objects at that velocity.


\section{Discussion}\label{sec:discuss}

\subsection{Kinematic Distance Discrimination}

\subsubsection{EMAFs as Distance Discriminators}\label{disc:discrim}

\begin{deluxetable*}{cccccccc}
  \tablecolumns{8}
  \tablewidth{0pc}
  \tabletypesize{\small}
  \tablecaption{Effect of Dust Temperature on KDA Resolution\label{table:td_kda}}
  \tablehead{
    \colhead{$T_d$} & \colhead{$N_{_\mathrm{WC}}$\tablenotemark{a}} & 
    \multicolumn{3}{c}{KDA Resolution\tablenotemark{b}} & \colhead{} & 
    \multicolumn{2}{c}{GRS Comparison} \\
    \cline{3-5} \cline{7-8}
    \colhead{(K)} & \colhead{} & \colhead{N} & \colhead{F} & \colhead{T} &
    \colhead{} & \colhead{N} & \colhead{Rate\tablenotemark{c}}
  }
  \startdata
  15 & 625 & 416 & 175 & 34 & & 218 & 77.5\% \\
  20 & 618 & 523 &  70 & 25 & & 213 & 91.9\% \\
  25 & 605 & 547 &  33 & 25 & & 198 & 91.9\%
  \enddata
  \tablenotetext{a}{Number of well-constrained distance estimates}
  \tablenotetext{b}{N = near; F = far; T = tangent point}
  \tablenotetext{c}{Distance matching success rate}
\end{deluxetable*}

The combination of millimeter-wave thermal dust emission observations with mid-infrared extinction is a powerful method for resolving the KDA for molecular cloud clumps.  By starting from a catalog of (sub-)millimeter sources, this method is not limited to mid-infrared-identified IRDCs (catalogs of which often have large minimum contrast).  We are therefore able to include low-column nearby sources as well as more distant objects.  The 93\% success rate compared to distances resolutions by the GRS team indicates that EMAFs can provide a powerful means for distance discrimination.  Additionally, BGPS objects placed at the far kinematic distance that agree with the GRS distance indicate that EMAFs are visible beyond the tangent point with sufficient backlighting.  In comparison with the HISA KDA-resolution technique employed by \citet{RomanDuval:2009}, only 4\% of BGPS objects were placed at the near kinematic distance, yet had no evidence of a HISA signature.  Application of additional prior DPDFs may help to resolve these discrepancies.

The mid-infrared contrast distributions (Fig.~\ref{fig:nf_hists}{\it c}) of objects on either side of \dtan\ clearly show that objects placed at or beyond the tangent point have lower collective contrast, and would likely not be included in catalogs of IRDCs.  These distributions are consistent with the notion that dark IRDCs ($C \gtrsim 0.2$) are nearby.  Since the matching rate between DPDF-derived distances and those of the GRS is nearly independent of mid-infrared contrast, the present method extends robust KDA resolution to EMAFs with lower contrast, roughly doubling the number of molecular cloud clumps for which well-constrained distances may be derived.

In addition to improving upon the axiom ``if IRDC then near'' for KDA resolution, this method automatically accounts for the profile of the 8-\micron\ intensity as a function of Galactic longitude (Fig.~\ref{fig:3d_ffore}).  The morphological matching process does not consider \dtan, and therefore offers a prior probability that is independent of the kinematic signature of a given object.  The \ffore\ limit of visibility for a molecular cloud clump is simply a function of optical depth (Equation \ref{eqn:cfs}); for instance, an object with $S_{_{1.1}} = 0.3$~Jy will have $C\geq0.05$ for \ffore~$\leq0.76$, but one with $S_{_{1.1}} = 0.1$~Jy will not meet this contrast threshold if \ffore\ exceeds 0.33.

Heightened star-formation activity, which produces excess 8-\micron\ emission in its immediate vicinity, does constitute a complicating factor in application of simple radiative transfer (Equation~\ref{eqn:rad_xfer}).  These regions strain the assumption of smooth, axisymmetric Galactic emission.  As an example, we analyzed the distance resolutions of objects in the W43 region.  W43 is defined here by $31\fdg5$~$\geq$ $\ell$~$\geq$ $29\fdg5$, $-0\fdg5$~$\geq$ $b$~$\geq0\fdg3$, and 80~\kms~$\geq$ \vlsr~$\geq$ 110~\kms\ \citep[as in][]{NguyenLuONg:2011}, and is marked by a white box in Figure~\ref{fig:lv_co}.  Of the 43 EMAF-selected BGPS sources with well-constrained distance estimates in this region, 9 are placed at or beyond the tangent distance by the DPDF method.  This is a slightly higher rate than the general sample, but is not significant.  The W43 objects are plotted as cyan squares in Figure~\ref{fig:sc_ratio}, and obey the empirical $S_{_{1.1}} / C = 2.5$ limit for near versus far distance discrimination (\S \ref{res:dist_helio}).  Only 8 objects could be associated with a GRS-identified \thco\ cloud (\S\ref{res:grs_comp}), and all but one have matching KDA resolutions; the outlier is one of the BGPS objects near the tangent point, where \dbar\ is used, causing the $>1$~kpc distance discrepancy.  Although the observed Galactic plane consists of clumpy emission atop a more smooth Galactic emission pattern, the simple model used here still returns consistent KDA resolutions, even in more active regions.

While the KDA resolutions for EMAF-selected BGPS objects compare favorably with previously-published distance estimates, they are still based upon the assumption of circular orbits about the Galactic center.  As highlighted by the case of CH$_3$OH maser G23.66-0.13, radial streaming motions can have a significant impact upon the derived kinematic distances.  \citet[][]{Anderson:2012} presents a detailed analysis of uncertainties involved with the use of kinematic distances in the presence of non-circular motions.  Future improvements in kinematic distance measurements will require a full three-dimensional vector model of Galactic motions.

Throughout this analysis we used $T_d = 20$~K as the temperature for converting BGPS flux densities into 8-\micron\ optical depths.  The effect of different assumed dust temperatures on KDA resolutions is not {\it a priori} predictable.  Generating a new set of DPDF$_\mathrm{emaf}$ based on different temperatures, however, is straightforward.  The results of KDA resolutions from the posterior DPDFs and GRS distance comparison statistics are shown in Table~\ref{table:td_kda} for the range of $T_d$ found by \citet{Battersby:2011}.  A warmer dust temperature pushes more objects to the near kinematic distance to compensate for a smaller derived $\tau_{_8}$.  Interestingly, the GRS success rate is unchanged for $T_d = 25$~K, but there are likely many unconsidered systematic effects at play.

The use of EMAFs as prior DPDFs for kinematic distance discrimination is directly applicable to all current and future (sub-)millimeter surveys of the Galactic plane.  The advantage of starting with a sample of continuum-identified molecular cloud clumps is that mid-infrared extinction may be used to resolve the KDA for many objects that may not be dark enough to be included in IRDC catalogs (\eg S06, PF09).

\subsubsection{Use of Different Distance Estimates}\label{disc:dist_est}

\begin{figure}[t]
        \centering
        \includegraphics[width=3.1in]{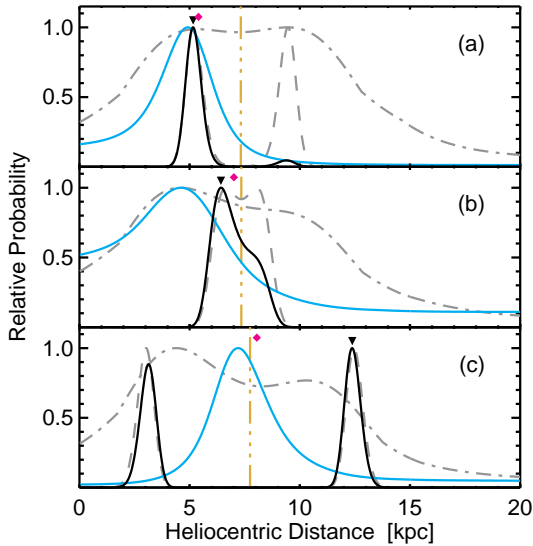}
        \caption{Example DPDFs for three possible cases.  Shown in each panel are DPDF$_\mathrm{kin}$ (dashed gray), DPDF$_\mathrm{emaf}$ (solid cyan), DPDF$_{\mathrm{H}_2}$ (dot-dashed gray), and posterior DPDF (solid black).  The vertical dot-dashed line in each panel marks the tangent distance along that line of sight.  The single-distance estimates are marked as black triangles (\dml) and magenta diamonds (\dbar).  Panel ({\it a}) represents a well-constrained KDA far from \dtan; ({\it b}) shows a source near \dtan; ({\it c}) illustrates a source with an unconstrained KDA resolution.  (A color version of this figure is available in the online journal.)}
        \label{fig:dpdf_example}
\end{figure}

The distance probability density function (DPDF) formalism encodes all information about distance determinations for molecular cloud clumps, including likely distance and uncertainty.  For some purposes, however, it is useful to have a single distance; \S \ref{dpdf:dist} describes two possible options.  In the use of the DPDFs produced here, it became apparent that different distance estimates were best applied to different situations.  Examples of these situations are shown in Figure~\ref{fig:dpdf_example} to illustrate the difficulties encountered in extracting a single distance from a DPDF.  For each  panel, the black triangle and magenta diamond mark \dml\ and \dbar, respectively.

The most common situation has well-separated kinematic distance peaks (\ie the molecular cloud clump is far from \dtan) and the probability ratio of the two peaks is also large (\ie \pchoose~$\geq 0.78$).  The example BGPS \#4484 (G030.629-00.029) is depicted in Fig.~\ref{fig:dpdf_example}{\it a}.  These objects fall under the ``well-constrained'' condition described in \S \ref{res:constr}, whereby the maximum-likelihood error bars encompass only one kinematic distance peak.  For this set of objects, \dml\ is a reasonable collapse of the DPDF into a single value (with uncertainty).  The second set of objects are those with kinematic distances within a kiloparsec of the tangent point.  The kinematic distance DPDF for these objects have a shallow saddle feature at \dtan, as seen in Fig.~\ref{fig:dpdf_example}{\it b} for BGPS \#4357 (G030.321+00.292).  Since the main peak of the posterior DPDF is not symmetric, \dml\ is not a robust reflection of the distance estimate.  Therefore, we recommend using \dbar\ for these sources (listed in Table~\ref{table:sources} for this class of object) to more accurately reflect the available distance information.  So long as the full-width of the error bars is less than 2.2~kpc (\S \ref{res:constr}), these sources are considered to have well-constrained distance estimates.  The final set of sources are those not meeting either of the above criteria, such as BGPS \#3352 (G024.533-00.182; Fig.~\ref{fig:dpdf_example}{\it c}).  The kinematic distance options are well-separated, but the DPDF$_\mathrm{emaf}$ does not place a strong discriminatory constraint.  If it is desirable to use the distances for these sources, we recommend Monte-Carlo sampling distances from the full DPDF in order to include all distance information about the source (see \S \ref{dpdf:physical}).  

Regardless of the method used, however, care should be taken to properly propagate the uncertainty in the distance placement.  If only sources in the first category are used, it is important to remember that matching success rate with GRS-derived distances was $\approx 92\%$.  This may be interpreted either as two sources in 25 were placed at the wrong distance or that there is a 92\% confidence in each of the distance placements.

\subsection{EMAF vs. IRDC}\label{disc:irdc}

In this paper, we introduce the nomenclature ``Eight-Micron Absorption Feature'' (EMAF) for dust-continuum-identified molecular cloud clump whose emission morphology matches an absorption feature in mid-infrared maps of the Galactic plane.  Many of these objects are quite dark ($C \gtrsim 0.2$) and are identified in IRDC catalogs.  To better understand the overlap between the EMAF and IRDC designations, we searched through both the S06 and PF09 catalogs to find the closest IRDC to the location of peak BGPS flux density.  A total of 361 (46\%) EMAF-selected BGPS sources lay within the semi-major axis distance of the centroid of a cloud from one or both of the catalogs.

\begin{figure}[t]
        \centering
        \includegraphics[width=3.1in]{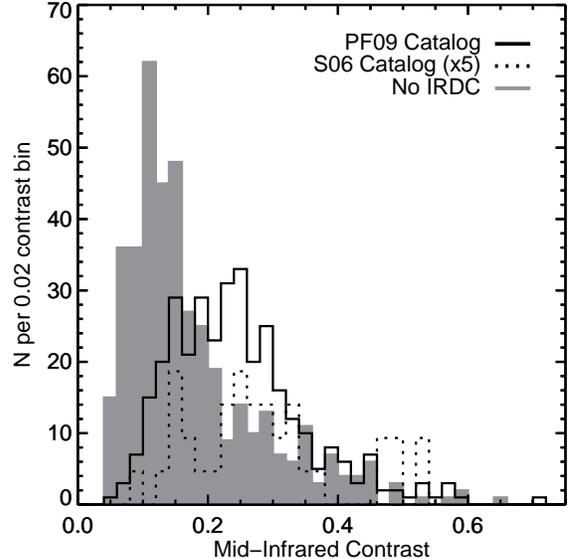}
        \caption{Histograms of EMAF contrast as a function of IRDC identification.  The black open histogram depicts objects associated with the PF09 catalog; black dotted shows those associated with S06 ($\times5$ for clarity); filled gray represents objects associated with neither catalog.}
        \label{fig:irdc_hist1}
\end{figure}

By definition, IRDCs have a large mid-infrared contrast, but EMAFs are selected from thermal dust emission catalogs.  As such, the two groups have different contrast distributions, as seen in the histograms of Figure~\ref{fig:irdc_hist1}.  BGPS sources that are associated with an object in the PF09 or S06 catalogs are plotted as black solid and dotted lines, respectively.  The filled gray histogram represents EMAFs not associated with any IRDC.  The bulk of the non-IRDC objects have $C\lesssim0.2$, once again confirming that the EMAF designation allows the use mid-infrared observations for KDA resolution for objects with low contrast.  Of particular interest are the higher-contrast ($C\gtrsim0.2$) BGPS sources which do not appear in either IRDC catalog (filled gray).  One striking example is the source G035.478-00.298 (BGPS \#5631), shown in the lower-right corner of Figure~\ref{fig:glimpse_ubc}{\it a}.  These objects suggest that it is easier to identify molecular cloud clumps from (sub-)millimeter data than to try to find intensity decrements in $\lambda = 8$~\micron\ images.

The EMAF-derived KDA resolutions of the IRDC objects place only a small fraction at the far kinematic distance (3\% and 7\% for S06 and PF09, respectively).  The measured EMAF contrast for such objects is generally $C\lesssim0.2$, reinforcing the notion that dark IRDCs are nearby.  Since the fractions of IRDCs placed beyond \dtan\ are comparable to the GRS distance mismatch rate (\S \ref{res:grs_comp}), these subsets are not significant.

\subsection{Implications for Galactic Structure}\label{disc:gal_struct}

\subsubsection{Galactic 8-\micron\ Emission}\label{disc:8um}

A quick glance at the GLIMPSE mosaics suggests that the Galactic distribution of 8-\micron\ light cannot be described simply by a smooth, diffuse model.  Distances derived using the assumption of smooth emission, however, compare favorably to those derived from 21-cm \HI\ absorption \citep[GRS;][]{RomanDuval:2009} and near-infrared extinction mapping \citep[NIREX;][]{Marshall:2009}.  These results suggest that Galactic 8-\micron\ emission may be primarily composed of a diffuse component punctuated by regions of active star formation.  The R12 model of mid- and far-infrared emission is a greatly simplified reflection of the Galaxy, neglecting to account for individual small-scale features.  Yet its match, spatially and spectrally, to existing Galactic plane observations indicates its power.  The model, therefore, provides a solid basis for our distance estimation technique based on using a broad distance prior to distinguish between kinematic distance peaks.

\begin{figure}[t]
        \centering
        \includegraphics[width=3.1in]{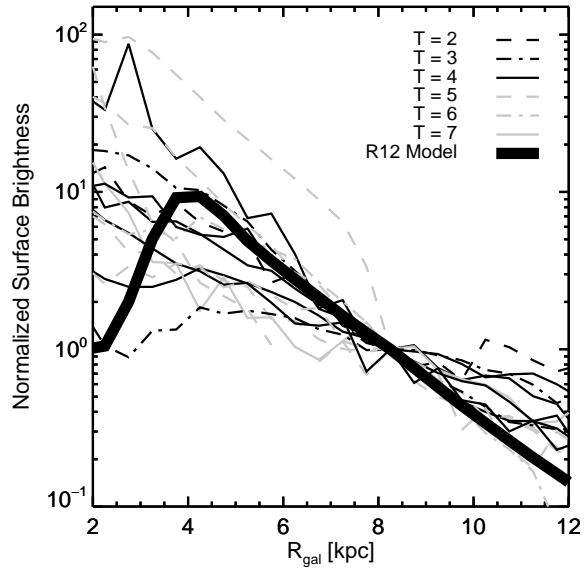}
	\caption{Azimuthally-averaged surface brightness ($\lambda = 8$~\micron) as a function of galactocentric radius.  A sample of 17 SINGS galaxies is plotted in various line styles according to Hubble stage ($T$), and are normalized to $R_\mathrm{gal} = 8.5$~kpc.  The thick black line depicts the surface brightness of the mid-infrared emission model (Fig.~\ref{fig:galmodel}).  The depression in the model profile at low $R_\mathrm{gal}$ is due to the dust hole in the R12 model (\S\ref{morph:model}).}
        \label{fig:radfield}
\end{figure}

While the R12 model was constructed to broadly match the multi-band observations of the Milky Way, it is useful to also compare it with external galaxies.  \spitzer\ observations of SINGS galaxies \citep{sings} yield a surface density of in-band emission.  Figure~\ref{fig:radfield} depicts the IRAC Band~4 surface brightness, normalized to $R_\mathrm{gal} = 8.5$~kpc for a collection of 17 SINGS galaxies, with galaxies represented by their Hubble stage ($T$)\footnote{Hubble stage is a continuous numerical representation of the Hubble type for a galaxy.  $T=0$ corresponds to an S0 galaxy, and the Milky Way (SBb-c) is $T \approx 4$ \citep[][p.155]{Binney:1998}.}.    To enable comparison, the R12 model was viewed externally (see Fig.~\ref{fig:galmodel}) and emission was annularly integrated, following the SINGS analysis; plotted as a thick black line.  The slope of the Milky Way model at $R_\mathrm{gal}\gtrsim 4$~kpc is comparable to the ensemble, but the drop in emission due to the dust hole carved out by R12 near the Galactic center does not seem to match extragalactic observations.  The model was designed to match observations from {\it within} the disk from the Sun's location, however, so discrepancies in the integrated 8-\micron\ emission profile in the inner $R_\mathrm{gal} \leq 3$~kpc are likely not relevant.

\subsubsection{Spiral Structure}\label{disc:spiral}

Recent large (sub-)millimeter Galactic plane surveys are making it possible to trace the spiral structure of the disk.  Comparing the well-constrained KDA resolutions of BGPS sources with an artist's conception of the Galaxy based on \spitzer\ data (Fig.~\ref{fig:galpos2}), glimmers of organization begin to appear.  While the regions that trigger the collapse of molecular cloud clumps are likely very localized along spiral density waves, features in the face-on map of the Galaxy derived from kinematic distances are quite smeared out.  Even small ($\approx 5$~\kms) peculiar motions can lead to $\approx 0.4$~kpc variations in heliocentric position, making it difficult to precisely trace the locations of spiral features.

Two major features suggest themselves from the data in Figure~\ref{fig:galpos2}: one near and one far.  First, the nearby collection of sources at $\ell \lesssim 30\degr$ seem to form part of a round feature that extends into the fourth quadrant.  This feature has been identified as both the Molecular Ring \citep[\cf][]{Dame:2001} and the Scutum-Centarus Arm \cite[\cf][]{Dobbs:2012}.  It is not possible to distinguish between these postulates in the northern plane; careful distance determinations for southern sources is required.  Mapping the exact location of where this feature meets the long Galactic bar also depends on choice of rotation curve (the \citealp{Clemens:1985} curve places this collection of sources at the tangent distance, whereas the \citealp{Reid:2009} curve does not), and may also be influenced by non-circular motions.  Parallax measurements to masers in this region will help establish a benchmark for the long Galactic bar, including position angle with respect to the Sun $-$ Galactic Center line.

The other feature to note in Figure~\ref{fig:galpos2} is the Sagittarius Arm beyond the tangent circle (smaller dotted circle).  These molecular cloud clumps are visible at the far kinematic distance because of backlighting provided by the Perseus Arm.  By the same token, BGPS sources in the Perseus Arm are not visible as EMAFs due to both the large amount of 8-\micron\ light in the foreground and the lack of any significant backlighting source.  We note a collection of sources around $\ell \sim 30\degr$ which appear at the far kinematic distance.  In the underlying image, there is a void between the Sagittarius Arm and the long Galactic bar.  The two possibilities, therefore, are the existence of an arm structure at that location or that those sources were improperly assigned the far kinematic distance.  The analysis of the W43 region, however, suggests that active regions do not significantly misplace molecular cloud clumps at the far kinematic distance, so distinguishing between the possibilities is unclear.


\section{Conclusions}\label{sec:conclusion}

We developed DPDFs as a new method for distance determinations to molecular cloud clumps in the Galactic plane.  Starting from a kinematic distance derived from molecular line observations as the likelihood, prior DPDFs may be applied in a Bayesian manner to resolve the KDA.  In this study, we used two external data sets as priors: mid-infrared absorption features, and the Galactic distribution of molecular gas.

The dust in molecular cloud clumps detected by (sub-)millimeter Galactic plane surveys should absorb mid-infrared light and be visible against the broad diffuse PAH emission near $\lambda = 8$~\micron.  Starting from the BGPS catalog of dust-continuum-identified molecular cloud clumps, we identified 770 EMAFs in the \spitzer/GLIMPSE mosaics.  EMAFs may be thought of as generalized IRDCs, and are characterized by their selection from (sub-)millimeter data.  With this collection of objects, simple radiative transfer arguments, and a model of Galactic mid-infrared stellar and dust emission, we developed a morphological matching scheme to compare dust emission and absorption.  When using the GLIMPSE mosaics to measure apparent absorption features, it is imperative to account for scattering of light within the IRAC camera.  This scattering, in concert with the instrumental calibration method, means that diffuse emission will appear brighter than it really is; bright emission will tend to fill in absorption features.  The scattered light changes the apparent contrast of absorption features, in addition to any derived properties (such as optical depth, mass, etc.).

Well-constrained KDA resolutions were obtained for 618 objects in this sample: 527 at the near kinematic distance, 31 at the tangent point, and 70 at the far kinematic distance.  To corroborate our distance discriminations, we used VLBI maser parallax measurements, KDA resolutions from the GRS, and near-infrared extinction distances as comparison sets.  Of the 13 objects associated with maser parallax measurements, none had a discrepant KDA resolution.  Distance comparisons with the GRS yielded a 93\% success rate nearly independent of mid-infrared contrast.  Comparison with the NIREX distances showed only one discrepant KDA resolution, with the remainder being within identified systematic effects.  These comparisons illustrate the validity of the present method, including the placement of some EMAFs at the far kinematic distance (approximately 12 distance-atched GRS sources are beyond \dtan).

Approximately half of the set of EMAFs are associated with an object from the IRDC catalogs of S06 and PF09.  Objects associated with IRDCs are mostly relatively dark $(C\gtrsim0.2)$; the remainder being largely low-contrast $(C\lesssim0.2)$.  Interestingly, there are a handful of moderately dark $(C\gtrsim0.3)$ EMAFs that are absent from these IRDC catalogs.  This suggests that it is perhaps easier to identify molecular cloud clumps first in (sub-)millimeter data, then investigate their mid-infrared properties.

KDA resolutions for EMAFs from the BGPS catalog reveal hints of Galactic structure.  Foremost, most detectable Galactic molecular cloud clumps are in the Molecular Ring / Scutum-Centarus Arm feature between the Sun and the Galactic center.  The Sagittarius Arm outside $\ell = 30\degr$ is suggested by a collection of EMAFs beyond the tangent point, visible due to backlighting from the more-distant Perseus Arm.

The derivation of DPDFs allows for probabilistic determination of distances to molecular cloud clumps across the Galactic plane.  By introducing the concept of an EMAF, we were able to use the mid-infrared GLIMPSE data to resolve the KDA for many more sources than is possible with extant catalogs of IRDCs.  Although this method applies only to $\sim10\%$ of the BGPS catalog, the DPDF framework allows for the incorporation of additional prior DPDFs to expand the number of molecular cloud clumps with well-constrained distance estimates.

\acknowledgments

The authors wish to thank N.~Halverson and J.~Kamenetzky for useful discussions, S.~Carey for help with understanding the scattering within the IRAC camera, and T. Robitaille for assistance with using {\sc Hyperion}.  This work was supported by the National Science Foundation through NSF grant AST-1008577.  The BGPS project was supported in part by NSF grant AST-0708403, and was performed at the Caltech Submillimeter Observatory (CSO), supported by NSF grants AST-0540882 and AST-0838261.  The CSO was operated by Caltech under contract from the NSF.  Support for the development of Bolocam was provided by NSF grants AST-9980846 and AST-0206158.  ER and SM are supported by a Discovery Grant from NSERC of Canada.  NJE is supported by NSF grant AST-1109116.  This work is based in part on observations made with the \spitzer\ Space Telescope, which is operated by JPL, Caltech under a contract with NASA.  This work utilized the Janus supercomputer, which is supported by the NSF (CNS-0821794) and the University of Colorado, Boulder.  Janus is a joint effort of CU-Boulder, CU-Denver and NCAR.  The GRS is a joint project of Boston University and Five College Radio Astronomy Observatory, funded by the NSF under grants AST-9800334, AST-0098562, AST-0100793, AST-0228993, \& AST-0507657.



\appendix

\section{Computing a Band-Averaged Dust Opacity: \spitzer\ IRAC Band~4}\label{app:dust_opacity}

The apparent dust opacity of extinction features in broadband images is related to the dust opacity as a function of frequency and the spectrum of the light being absorbed.  The IRAC Band~4 bandpass includes several distinct emission features from PAH molecules, as well as the complex behavior of the dust opacity near the 10-\micron\ silicate feature.  In this appendix, we derive a band-averaged dust opacity $\kband$ for use with the GLIMPSE mosaics.

Given the simple radiative transfer model of Equation~(\ref{eqn:rad_xfer}), the intensity transmitted through a dust cloud is
\beqn\label{eqn:i_trans}
I_\mathrm{trans} = I_\mathrm{back}~e^{-\tauband}~.
\eeqn
where $I_\mathrm{back}$ is the background light (from the cloud to large heliocentric distance), and $I_\mathrm{trans}$ is the transmitted light exiting the cloud on the near side (\ie not including emission between the cloud and the observer).  The band-averaged optical depth is related to the apparent dust opacity by $\tauband = \Sigma~\kband$, where $\Sigma$ is the mass surface density of dust.  The band-averaged dust opacity is therefore
\beqn\label{eqn:kband}
\kband = - \frac{1}{\Sigma} \ln \left(\frac{I_\mathrm{trans}}{I_\mathrm{back}} \right)~.
\eeqn

The ratio of intensities is computed from the band average over each quantity,
\begin{eqnarray}\label{eqn:i_ratio}
\frac{I_\mathrm{trans}}{I_\mathrm{back}} &=& \frac{\langle I_\mathrm{back,\nu}\ e^{-\tau_\nu}\rangle_{_\mathrm{band}}}{\langle I_\mathrm{back,\nu} \rangle_{_\mathrm{band}}} \nonumber \\
  &=& \frac{\int R_{_\mathrm{band}}(\nu)\ I_\mathrm{back,\nu}\ e^{-\tau_\nu}\ d\nu }{\int R_{_\mathrm{band}}(\nu)\ I_\mathrm{back,\nu}\ d\nu}\ ,
\end{eqnarray}
where the $\nu$ subscript denotes that quantity as a function of frequency, and $R_{_\mathrm{band}}(\nu)$ is the relative frequency response per unit power for the instrument.  Since both averages are over the same response bandpass, the usual normalization terms cancel.

The typical radius of an IRDC is small \citep[$\sim1$~pc;][]{Rathborne:2006} compared to the accumulated path length ($D$) for the diffuse background (several kiloparsecs), so the intensity $I_\mathrm{back,\nu} = \int j_\nu\ ds$ may be approximated by $I_\mathrm{back,\nu} = j_\nu~D$, where $j_\nu$ is the emission coefficient.  Additionally, the relative response per unit power, $R_{_\mathrm{band}}(\nu)$, is proportional to $(1/h\nu)\ S_{_\mathrm{band}}(\nu)$, where $S_{_\mathrm{band}}(\nu)$ is the relative response per photon\footnote{Obtained from \texttt{http://irsa.ipac.caltech.edu/data/SPITZER/docs/irac/calibrationfiles/spectralresponse}.}.  Canceling frequency-independent quantities and inserting the intensity ratio into Equation~(\ref{eqn:kband}) yields the desired band-averaged dust opacity,
\beqn\label{eqn:kband_total}
\kband = - \ln \left[\frac{\int \nu^{-1}\ S_{_\mathrm{band}}(\nu)\ j_\nu\ e^{-\kappa_\nu}\ d\nu }{\int \nu^{-1}\ S_{_\mathrm{band}}(\nu)\ j_\nu\ d\nu} \right]~.
\eeqn

For the emission spectra ($j_\nu$), we used the dust emission models of \citet{Draine:2007}, which contain a mixture of grain sizes in addition to a variable PAH mass fraction ($q_{_\mathrm{PAH}}$).  The various emission spectra were derived by irradiating the dust with starlight intensity fields having a tunable minimum value $U_\mathrm{min}$ relative to the local interstellar radiation field ($U=1$).

\begin{deluxetable*}{lccccc}
  \tablecolumns{6}
  \tablewidth{0pc}
  \tabletypesize{\small}
  \tablecaption{Computed $\kband$ for IRAC Band~4\label{table:kband}}
  \tablehead{
    \multicolumn{2}{c}{Dust Emission\tablenotemark{a}} & \colhead{} &\multicolumn{3}{c}{$\kband$} \\
    \cline{1-2} \cline{4-6}
    \colhead{$q_{_\mathrm{PAH}}$} & \colhead{$U_\mathrm{min}$} & \colhead{} & \colhead{OH5\tablenotemark{b}} & \colhead{WD01-3.1\tablenotemark{c}} & \colhead{WD01-5.5\tablenotemark{d}}\\
    \colhead{(\%)} & \colhead{} & \colhead{} & \multicolumn{3}{c}{(cm$^2$ g$^{-1}$)}
  }
  \startdata
0.47...... & 0.1 &  & 1175 & 858 & 909 \\
 & 1.0 &  & 1177 & 861 & 911 \\
 & 10.0 &  & 1178 & 864 & 915 \\
\hline
2.50...... & 0.1 &  & 1157 & 796 & 843 \\
 & 1.0 &  & 1158 & 800 & 847 \\
 & 10.0 &  & 1160 & 804 & 852 \\
\hline
4.58...... & 0.1 &  & 1150 & 777 & 823 \\
 & 1.0 &  & 1151 & 779 & 825 \\
 & 10.0 &  & 1152 & 782 & 828
  \enddata
\tablenotetext{a}{Dust emission model from \citet{Draine:2007}}
\tablenotetext{b}{Extinction from \citet[][Table 1, Column 5]{Ossenkopf:1994}}
\tablenotetext{c}{$R_V = 3.1$ extinction from WD01, with updated normalizations from \citet{Draine:2003a}}
\tablenotetext{d}{$R_V = 5.5$, Case~A, extinction from WD01, with updated normalizations from \citet{Draine:2003a}}
\end{deluxetable*}

The choice of dust opacity model ($\kappa_\nu$) has a nontrivial effect on the derived band-averaged opacity.  Three different models were analyzed, and are shown in Table~\ref{table:kband}.  First is the OH5 model \citep[dust grains with thin ice mantles, coagulating at 10$^6$~cm$^{-3}$ for 10$^5$~years;][]{Ossenkopf:1994} used for determining the dust opacity at $\lambda = 1.1$~mm for the BGPS.  The remaining models were presented in \citet[][hereafter WD01]{Weingartner:2001}, with updated normalizations given by \citet{Draine:2003a}\footnote{\texttt{http://www.astro.princeton.edu/$\sim$draine/dust/dustmix.html}}.  The second of the three models is the $R_V = 3.1$ Milky Way model, tried even though this value of the color excess per magnitude extinction is consistent with the diffuse ISM and does not hold for regions of dense gas.  The final model utilizes case~A for $R_V = 5.5$ (consistent with observations of molecular clouds), which sought to minimize the extinction differences between observation and model, while also including a penalty term to keep dust grain volume from exceeding abundance/depletion limits (WD01).  For each dust model, $\kband$ was computed for three values each of $q_{_\mathrm{PAH}}$ and $U_\mathrm{min}$ (Table~\ref{table:kband}).

The minimum value of the starlight intensity field has very little effect on the band-averaged dust opacity, meaning that the derived $\kband$ are valid for a wide range of environments.  An order of magnitude change in the assumed PAH mass fraction causes only a 2\% difference in the derived opacity for the OH5 model, but the spread is 10\% for the others.  \citet{Draine:2007} cite $q_{_\mathrm{PAH}} = 4.58\%$ as best matching observations of the Milky Way.

The OH5 model is the preferred description of dust at (sub-)millimeter wavelengths \citep[\cf][]{Rathborne:2006,Schuller:2009,Aguirre:2011}, and has been used by previous studies for estimating $\kband$ for IRAC Band~4 images \citep{Butler:2009,Battersby:2010}.  That model was computed from theory for coagulated grains (aggregates of smaller particles, with some voids) surrounded by an ice mantle.  In contrast, the WD01 dust models utilized simple geometry (PAH molecules for very small grains, and graphite and olivine spheres for larger grains) and sought to fit a dust size distribution to parameterized observed extinction.

The connection to observed extinction in the infrared led us to choose the WD01 $R_V = 5.5$ model for this work.  Following \citet{Draine:2007}, we used $q_{_\mathrm{PAH}} = 4.58\%$ and $U_\mathrm{min} = 1.0$ to compute $\kappa_8 = \kband = 825$~cm$^2$~g$^{-1}$ of dust.  For comparison, the corresponding value from the OH5 model yields $\kappa_8 = 1167$~cm$^2$~g$^{-1}$, a $\approx40\%$ difference.  We note that the preferred WD01 model predicts a BGPS dust opacity $\kappa_{_{1.1}} = 0.272$~cm$^2$~g$^{-1}$ of dust, approximately one quarter the value from OH5.

\section{\spitzer\ IRAC Scattering Correction Factors}\label{app:scatter}

The IRAC camera on the \spitzer\ Space Telescope suffers from internal scattering within the detector arrays, particularly Bands 3 and 4.  The scattering is such that a fraction of the incident light on a pixel is distributed throughout the entire array (\citealp{Reach:2005}; IRAC Instrument Handbook).  Image frames are converted into physical units (MJy~sr$^{-1}$) using point-source calibration data; point-source aperture photometry is therefore accurate because the calibration takes into account the light scattered out of the aperture and into blank sky pixels.  Observed extended emission, however, has light from other areas of the array scattered {\it into} each pixel as well, and so will appear brighter than it really is given the point-source calibration.  For the broad, diffuse emission of the Galactic plane in IRAC Band~4, there is a roughly constant positive offset of the measured intensity in each frame.

For absorption features in the Band~4 images (EMAFs), however, the scattering cannot be corrected for by simple multiplicative aperture corrections.  Because bright emission from surrounding regions is scattered into an EMAF, it will have a lower apparent contrast.  To correct for this effect, an estimate of the scattered light in a frame must be {\it subtracted} from each frame (S. Carey 2010, private communication), as was done in this study for a pixel-by-pixel comparison between GLIMPSE and synthetic 8-\micron\ images.

Quantities such as contrast and optical depth for EMAFs may be derived from the GLIMPSE mosaics (\eg \citealp{Butler:2009}; PF09), but a scattering correction must be applied \citep[\cf][]{Battersby:2010}.  Because careful subtraction of scattered light is not always necessary for a given application, correction factors may be derived for quantities measured directly from the IRAC Band~4 data.  In this appendix, we derive correction factors for mid-infrared contrast ($C$), 8-\micron\ optical depth ($\tau_{_8}$), and foreground fraction of 8-\micron\ emission (\ffore).

For regions of broad diffuse emission punctuated by dark clouds, the observed intensities $I_0$ and $I_1$ of the background and EMAF, respectively, are related to the actual intensities $S_0$ and $S_1$ by
\begin{eqnarray}
I_0 &=& S_0 + X~,~\mathrm{and} \label{eqn:i0} \\
I_1 &=& S_1 + X~, \label{eqn:i1}
\end{eqnarray}
where $X = \xi~S_0$ is the amount of scattered light, approximated by the fraction $\xi = (1-0.737) = 0.263$ of incident diffuse light scattered throughout the array, and 0.737 is the infinite-aperture correction for Band~4 from \citet{Reach:2005}.  Rearranging to compute the true intensities from observed quantities yields
\begin{eqnarray}
S_0 &=& \frac{I_0}{(1+\xi)}~,~\mathrm{and}\label{eqn:s0} \\
S_1 &=& I_1 - I_0\frac{\xi}{(1+\xi)}~. \label{eqn:s1}
\end{eqnarray}
While the subtractive correction (Equation \ref{eqn:i0}) for the diffuse background is equivalent to a multiplicative correction (Equation \ref{eqn:s0}), correcting for the intensity within the EMAF is more complicated.  To compute the true contrast of an EMAF, we begin with Equation (\ref{eqn:contrast}):
\beqn\label{eqn:true_c}
C_{_\mathrm{true}} = 1 - \frac{S_1}{S_0}~.
\eeqn
Inserting Equations (\ref{eqn:s0}) and (\ref{eqn:s1}), the true contrast becomes
\begin{eqnarray}
C_{_\mathrm{true}} &=& \left(1 - \frac{I_1}{I_0} \right) \left( 1 + \xi \right) \nonumber \\
 &=& C_{_\mathrm{meas}} \left(1 + \xi \right)~,
\end{eqnarray}
where $C_{_\mathrm{meas}}$ is the quantity measured directly from the GLIMPSE images.  The measured contrast will be {\it smaller} than reality, leading to an underestimation of optical depth and other quantities.

With a measured EMAF contrast, it is possible to estimate the optical depth of a cloud or the foreground fraction of diffuse emission, given an assumption about the other.  Equation (\ref{eqn:cfs}) may be rearranged to solve for either quantity in terms of the other.  The optical depth (from which follows surface mass density) is given by
\begin{eqnarray}
\tau_{_\mathrm{8,true}} &=& -\ln \left[ 1 - \frac{C_{_\mathrm{true}}}{1 - f_\mathrm{fore}} \right] \nonumber \\
 &=& -\ln \left[ 1 - \frac{C_{_\mathrm{meas}} \left(1 + \xi \right)}{1 - f_\mathrm{fore}} \right]~,
\end{eqnarray}
assuming a model that yields \ffore\ \citep[as in][]{Butler:2009}.  The true value of $\tau_{_8}$ will be larger by up to a factor of two for $C_{_\mathrm{meas}} \lesssim 0.5$.  If, instead, the foreground fraction is desired given an external estimate of $\tau_{_8}$ (such as from (sub-)millimeter thermal dust continuum data; as in PF09), the true value is given by
\begin{eqnarray}
f_\mathrm{fore,true} &=& 1 - \frac{C_{_\mathrm{meas}} \left(1 + \xi \right)}{1 - e^{-\tau_{_8}}} \nonumber \\
 &=& \left( 1 - \frac{C_{_\mathrm{meas}}}{1 - e^{-\tau_{_8}}} \right)\left(1 + \xi \right) - \xi \nonumber \\
 &=& f_\mathrm{fore,meas} \left(1 + \xi \right) - \xi~.
\end{eqnarray}
The actual foreground fraction will be smaller than that measured directly from GLIMPSE images, with the difference becoming less at large \ffore.  Any $f_\mathrm{fore,meas} \lesssim 0.2$ maps to zero true foreground fraction, as negative values are not physical; such values arise from uncertainty in $C$ and the derivation of $\tau_{_8}$ from (sub-)millimeter data.

\section{The Vertical Solar Offset and Converting \lbd\ into \rphiz}\label{app:coord_conv}

Deriving Galactocentric positions of objects in the Milky Way requires a coordinate transformation of the triad \lbd, where \dsun\ is the heliocentric distance along the line of sight toward ($\ell,b$).  The Galactic coordinate system was defined assuming the Sun is at the midplane of the disk \citep{Blaauw:1960}, but more recent studies have measured a vertical solar offset of $\approx 25$~pc above the midplane \citep{Humphreys:1995,Juric:2008}.  Since the vertical scale height of the molecular gas layer in the disk is small \citep[HWHM~$\approx 60$~pc;][]{Bronfman:1988}, neglecting to account for the solar offset may introduce a systematic bias in the derived vertical distributions of components of the Galactic disk.

The coordinate transformation is done in cartesian coordinates.  First the triad \lbd\ is converted into local cartesian coordinates, where the $x$-axis is directed along the Sun $-$ Galactic Center line, and $z$ points north out of the plane,
\beqn\label{eqn:local_coord}
\left( \begin{array}{c} x_1 \\ y_1 \\ z_1  \end{array} \right) =
\left( \begin{array}{c} d_{_\sun}\ \cos l\ \cos b \\ d_{_\sun}\ \sin l\ \cos b \\ d_{_\sun}\ \sin b \end{array} \right)~.
\eeqn
The local coordinates are transformed to the Galactocentric frame by (1) rotation by 180\degr\ in the $x-y$ plane to place the $+x$-axis pointing away from the GC, (2) translation of the coordinate axes to place the origin at the GC, and finally (3) rotation by the angle $\theta$ in the $x-z$ plane to place the $+x$-axis along the Galactic midplane (rather than along the Sun $-$ Galactic Center line).  The transformation may be written as
\beqn{
\left( \begin{array}{c} x_\mathrm{gal} \\ y_\mathrm{gal} \\ z_\mathrm{gal} \\ 1  \end{array} \right) =
\left( \begin{array}{cccc} \cos \theta & 0 & -\sin \theta & 0 \\
                           0 & 1 & 0 & 0 \\
                           \sin \theta & 0 & \cos \theta & 0 \\
                           0 & 0 & 0 & 1 \end{array} \right) 
\left( \begin{array}{cccc} 1 & 0 & 0 & R_0 \\
                           0 & 1 & 0 & 0 \\
                           0 & 0 & 1 & 0 \\
                           0 & 0 & 0 & 1 \end{array} \right) 
\left( \begin{array}{cccc} -1 & 0 & 0 & 0 \\
                           0 & -1 & 0 & 0 \\
                           0 & 0 & 1 & 0 \\
                           0 & 0 & 0 & 1 \end{array} \right) 
\left( \begin{array}{c} x_1 \\ y_1 \\ z_1 \\ 1 \end{array} \right) }\ ,
\eeqn
where the lateral translation requires an augmented (affine translation) matrix.  The resulting Galactocentric cartesian coordinates are
\beqn\label{eqn:xyz_gal}
\left( \begin{array}{c} x_\mathrm{gal} \\ y_\mathrm{gal} \\ z_\mathrm{gal}  \end{array} \right) =
\left( \begin{array}{c} 
R_0\ \cos \theta - d_{_\sun}\ (\cos l\ \cos b\ \cos \theta + \sin b\ \sin \theta) \\
-d_{_\sun}\ \sin l\ \cos b \\
R_0\ \sin \theta - d_{_\sun}\ (\cos l\ \cos b\ \sin \theta - \sin b\ \cos \theta)
\end{array} \right) \ .
\eeqn
The rotation angle $\theta = \sin^{-1} (z_0/ R_0)$, where $z_0 = 25$~pc, corrects for the Sun's vertical displacement above the midplane.  Galactocentric positions in the cylindrical coordinates ($R_\mathrm{gal},\phi,z$) may be extracted from Eqn.~(\ref{eqn:xyz_gal}) in the usual manner.  The rotation by $\theta$ is most important for the derived $z_\mathrm{gal}$, and has negligible effect on $R_\mathrm{gal}$ and $\phi$.

~

\end{document}